\documentclass[]{cjp}

\usepackage{graphicx}

\volyear{}[]{}
\received{}
\accepted{}

\begin{document}

\title{\bf Spectra of phosphorus ions for astrophysical modeling:
P~I - P~XV 
}
\author{Sultana N. Nahar} 
\address{Department of Astronomy, The Ohio State University, Columbus, OH 
43210, USA}
 \AddressNote{The authors are honored to participate in the dedication
issue for Prof. Ravi Rau who made pioneering contributions to photon-atom
interaction, and spectroscopy.}
\author{Bilal Shafique} 
\address{Astronomy, The Ohio State University, Columbus, OH 
43210, USA} 
\AddressNote{Current: Department of Physics, University of Azad Jammu 
$\&$ Kashmir, Muzaffarabad 13100, Pakistan}
\shortauthor{Nahar and Shafique.}

\maketitle

\begin{abstract}
Phosphorus (P), a basic element of life, has been a least studied element 
due to its poor presence in astrophysical spectra. However, search for the 
P lines has increased considerably with discoveries of exoplanets and are 
being detected by high resolution and sophisticated astronomical 
observatories, e.g. James Webb Space Telescope (JWST). JWST may provide 
a clue for life with detection of P in its infrared (IR) region. 
Identification of the element and analysis of the observed spectra 
will require high accuracy data for atomic processes that produces lines 
and their predicted features. 
The present study focuses on these needs and reports systematically regions 
of wavelengths, from x-ray to IR, that show prominent lines by the 15 
individual ionization stages of phosphorus, P I - P XV for the first time.
We present large amount of relevant atomic data for energies, transition 
parameters, and lifetimes obtained in relativistic Breit-Pauli 
approximation using the R-matrix method and atomic structure program 
SUPERSTRUCTURE. Our spectral features for the fifteen ions, P I - P XV,
predict strengths of lines in various wavelength regions.
They show dominance of P~I and P~II in the infrared region and other ions in 
the ultraviolet and optical regions often stretching to IR in the continuum. 
For determination of accuracy, we have made extensive comparisons of 
our atomic data with available experimental and theoretical values. 
Based on these, our results and features are expected to provide precise
plasma diagnostics and astrophysical modeling. 
\end{abstract}

\medskip


\date{today}

\section{Introduction}

Chemical elements, such as, carbon, nitrogen, oxygen, phosphorus, sulfur
form the basis of human life on the earth. Phosphorus (P) is contained 
in DNA-RNA and acts as an energy carrier, and hence plays a key role as 
a biosignature element. It is abundant in the solar system but lacks 
in space. The lack of P is often linked to not finding any life form 
outside the earth. Hence, with discoveries of exoplanets, phosphorus 
is holding a special importance to astrobiology as its relative 
abundance in a planetary host star will increase the prospects for 
life on its planet \cite{hetal20}. 
Search of its lines in the exoplanetary atmosphere has increased 
considerably.  James Webb Space Telescope (JWST) is expected to obtain 
high-resolution spectra in the infrared (IR) region of 0.6 - 28.3 
$\mu$m which includes a few ionization stages of phosphorus. Phosphorus 
has been  detected in a number of astronomical objects, such as, in 
damp Galaxies \cite{mol2001}, \cite{wel2001}, in nebular environments 
of supernova remnants of Cassiopeia \cite{ketal13}.
Lines of  P~II and P~III have been found in the high resolution 
spectra of hot OB stars under a project called ASTRO-2 (J. Hillier, 
private communication 2015).  However, poor abundance caused limited 
number of investigations of the element and hence a limited amount 
of atomic data available for spectral analysis of various ionization 
stages of P. 

Existing data of P ions are largely for energies for all ionization 
stages from P I - P XIV measured by Martin et al \cite{metal85} and
for P XV measured by Yerokhin and Shabaev \cite{ys15}, and Erickson
\cite{erickson77}. These energies are available in the compiled tables
at the website of National Institute of Standards and Technology (NIST) 
\cite{nist}. Literature search shows that the other atomic data available
are mainly for radiative transitions in a limited manner and radiative
lifetimes of levels measured using beam-foil, fluorescence detection
technique, and in storage ring and cyclotron experimental setups.
Only a few studies have been done for other atomic processes, such as,
photoionization, electron-ion recombination, electron impact excitation, 
for astrophysical plasma modeling applications. 
A brief summary of the past studies on the atomic processes for each 
P ion is given below.

{\bf P I:} NIST \cite{nist} compiled table reports transition probabilities 
calculated by Lawrence \cite{law67}, Czyzak and Krueger \cite{czy63}.
Energies for fine structure levels, radiative decay rates for both
allowed (E1) and forbidden (M1, E2, M2, and E3) transitions, and
lifetimes for a number of levels of neutral phosphorus (P-I) along with
other ions in isoelectronic sequences from sodium (Na) through Argon
(Ar) were calculated by Zatsarinny and Frose-Fischer \cite{zf2002} and
Frose-Fischer \cite{fisc2006}.
Lifetimes for the 21 excited states of phosphorus ions (P I through
P V) were measured by Curtis et al \cite{curtis71} 
from the spectra of phosphorus in the UV region (600 - 2200){\AA}
using beam-foil technique. Berzinish et al \cite{berz1997} measured
lifetimes of P~I using fluorescence detection method.

{\bf P II:} Transitions in P~II were studied by several investigators 
\cite{savage66,fuhr98,wiese69,czy63}.
The calculations for P II transitions were carried out using the MCDF   
atomic code developed and revised by Desclaux \cite{desc75}. Later Huang 
\cite{huang85}, Brown et al \cite{brown18} reported calculated energies 
and transition probabilities of P-II. Curtis et al \cite{curtis71} and
Brown et al \cite{brown18} measured lifetimes of P II.
Photoionization cross sections of many excited levels going up to n=10 
of P II were obtained using relativistic Breit-Pauli R-matrix (BPRM) 
method by Nahar \cite{snn17}.
Photoionization cross sections of the ground and low lying excited levels
of P II were benchmarked with experiment carried out at the Advanced Light
Source at LBNL by Nahar et al \cite{setal17}. Nahar \cite{snn17b} studied  
electron-ion recombination of P II using the unified method of Nahar and 
Pradhan \cite{np92,np94}. All atomic data for photoionization and electron-ion
recombination are available electronically at the NORAD-Atomic-Data 
database \cite{norad}.

{\bf P III:} The
transitions in P~III were studied by Wiese et al \cite{wiese69}, Fuhr
et al. \cite{fuhr98}, Huang \cite{huang86}, Naqvi \cite{naqvi51}.  
Naqvi calculated a single M1 forbidden transition. 
Huang \cite{huang86} reported energies and transition probabilities
of P-III along with other Si-like ions. Measurement of lifetimes of 
phosphorus ions by Curtis et al \cite{curtis71} included those of P III.
Measured spectra of photoionization cross sections for P III and P IV
for the ground and a few low lying levels were reported by Hernándeza
et al \cite{hetal15} and the measured resonances were identified 
theoretically by Gning et al \cite{getal19}.
The electron impact excitation collision strength and rates for P III
obtained using BPRM method by Naghma et al \cite{netal18}.


{\bf P~IV} Transitions i P~IV were studied by Zare \cite{zare67}, Crossly 
and Dalgarno \cite{cross65}. Naqvi \cite{naqvi51} calculated M1 transitions 
among the levels of $3sp3(^{1,3}P^o)$.
Lin et al \cite{lin78} used the semi-empirical model potential method 
to calculate the excitation energies and magnetic quadrupole (M2) 
transition rates for Na-, Be-, and Mg-like ions (including P-IV).
Godefroid et al \cite{gode85} measured and calculated the relative 
intensities of
the E2 transitions of P IV and P V.
Measured lifetimes of various levels of phosphorus ions have been 
reported by Curtis 
et al \cite{curtis71} which included those of P IV, Brown \cite{brown18}, 
Alkhayat \cite{khayat19}, Van Der Westhuizen et al \cite{west89},
Maio \cite{maio76}. 
Photoionization cross sections for P~IV were measured by Hermandeza et al
\cite{hetal15} and resonances calculated by Gning et al \cite{getal19}.

{\bf P~V:}
Transitions in P~V were studied by several investigators 
\cite{gode85,john96,cross65,wiese69}.
 Godefroid et al \cite{gode85} measured and calculated 
the relative intensities of the E2 transitions of P IV and P V.
Lifetimes of P~V were measured by Curtis et al \cite{curtis71}  and 
Maio \cite{maio76}.
 

{\bf P~VI:}
Earlier calculations for a limited transitions in P~IV were carried out by
Kastner et al \cite{kastner67}.
Hibbert \cite{hibb93} reported a large set of atomic data  
for level energies, oscillator strengths, lifetimes for neon-like ions
(Ne-I through Kr XXVII) which included P-VI  using the general-configuration 
interaction code, CIV3, which incorporates the optimized orbitals and a
modified Breit-Pauli approximation. 
Zhu et al \cite{zhu2016} used GRASPVU atomic code package, a modified version
of the general-purpose relativistic atomic structure package GRASP92, and
computed transition wavelengths, transition probabilities, line
strengths and the oscillator strengths for nine charged states of phosphorus
(P-VI through P-XIV).


{\bf P VII:}
The earlier work on the radiative transitions in P~VII include those of
Cohen and Dalgarno \cite{cohen64}, Naqvi \cite{naqvi51},
Aggarwal \cite{agg2019}. Arrarwal reported theoretical calculations of level 
energies for 198 levels, transition probabilities, and radiative lifetimes 
for F-like ions ($12 \leq Z \leq 23$) which included P VII. He used 
Flexible Atomic Code (FAC) which uses the relativistic approached based 
on Dirac equation for level energies and the general-purpose relativistic 
atomic structure package (GRASP) to calculate the radiative transition 
rates and lifetimes for the electric dipole and quadrupole transitions
(E1 and E2) and the magnetic dipole and quadrupole transitions (M1 and M2).
Cheng et al \cite{cheng79} reported transition probabilities for electric 
dipole (E1), quadrupole (E2), and magnetic dipole transitions for Li 
through F ioselectronic sequences of various ions including P-VII through 
P-XIV using MCDF method.
Zhu et al's \cite{zhu2016} calculations included the energies and transition 
parameters for P VII.


{\bf P VIII:}
Transitions in P VIII have been studied by Cohen and Dalgarno \cite{cohen64},
Cheng et al \cite{cheng79}, Naqvi \cite{cheng79}, Malville and
Berger \cite{malv65}, Zhu et al \cite{zhu2016}. 
Tr\"{a}bert et al \cite{trabert12} used Test Storage Ring (TSR) 
set-up to measure lifetimes of $2s^22p^2~^1D_{2}$, 
$2s^22p^3~^2P^o_{1/2,3/2}$, and
$2s^22p^4~ ^1D_{2}$ levels of various ions (Si through S) including
P-VIII, P-IX, and P-X.

{\bf P IX:}
Transitions for P~IX have been studied by Cheng et al \cite{cheng79}, 
Zhu et al \cite{zhu2016}, Cohen and Dalgarno \cite{cohen64}, Naqvi
\cite{cheng79}. 
Tr\"{a}bert et al \cite{trabert12}  used TSR set-up to measure lifetimes 
of levels of P-IX. 


{\bf P X:}
Energies and transition parameters for P~IX were calculated by Cheng 
et al \cite{cheng79}, Zhu et al \cite{zhu2016},
Cohen and Dalgarno \cite{cohen64}, Malville and Berger \cite{malv65},
Froese-Fischer \cite{froese66}, Naqvi \cite{naqvi51} reported
energies and transition parameters for P X.
Tr\"{a}bert et al \cite{trabert12} 
reported lifetimes of levels of P-X.


{\bf P XI:}
Cheng et al \cite{cheng79}, Zhu et al \cite{zhu2016} reported energies
and transitions in P~XI.  Cohen and Dalgarno \cite{cohen64}, Naqvi 
\cite{naqvi51}, Wiese et al. \cite{wiese69} also reported study of 
transitions in P~XI.


{\bf P XII:}
Energies and transitions of P~XII were reported by Cheng et al 
\cite{cheng79}, Zhu et al \cite{zhu2016}, Naqvi \cite{naqvi51}, Garstrang 
and Shamey \cite{garstang67}, Cohen and Dalgarno \cite{cohen64}, and 
Naqvi and Victor \cite{naqvi64}.
From the analysis of decay curves obtained from Beam-Foil excitation
Tr\"{a}bert and Heckmann \cite{trabert80a,trabert80b} deduced the lifetimes 
in the EUV region (10\AA - 550\AA) for 
P~XI.

{\bf P XIII:}
Transitions have been calculated by investigators \cite{zhu2016,cheng79,
wiese69,cohen64}. 
Experimental radiative decay rates of transitions in Li- and He-like P-ions
have been reported by Deschepper et al \cite{desc82} 
using Doppler-tuned x-ray spectrometer for the acquisition of highly resolved
spectra in x-ray region.


{\bf P XIV:}
Cheng et al \cite{cheng79}, Zhu et al \cite{zhu2016} reported the energies
and transition parameters for P XIV.
As mentioned above, measured radiative decay rates of transitions of He-like
P-ions have been reported by Deschepper et al \cite{desc82}, Wiese et al 
\cite{wiese69}, Cohen and Dalgarno \cite{cohen64}, Drake \cite{drake79},
and Lin et al \cite{lin1977}. 

{\bf P XV:}
The one-photon transition probabilities for H-like ions with  nuclear 
charges 1 $\leq$ Z $\leq$ 100 (including P-XV) using the relativistic 
quantum theory were reported by Popov and Maiorova \cite{popov2017}. 

The objectives of the present work is to obtain large set of energy 
levels and the transitions among them, and implement them to  study 
and predict the domain 
of prominent spectral features in broad wavelength ranges, from x-ray to
infrared (IR), for all 15 ionization stages of phosphorus, P I - P XV.
The present study is the first systematic study for a
relatively complete spectral features of all ionization stages of 
phosphorus.
Detection of features can lead to determination of presence of the element 
in astronomical objects and identify the ionization stages for modeling.
We have considered transitions from orbital $1s$ to going up highly
excited states. 
There are other atomic processes in addtion to photo-excitation that 
contribute to the formation of a spectrum.
For example, photoionization enhances the background and strengthens the
peaks of the spectral lines and may even appear as absorption lines in 
the continuum. Electron-ion recombination and electron-impact excitation 
also introduce lines that are used for diagnostics of the plasma. But a 
spectrum and its features are mainly created by photo-excitation. 

We have employed the relativistic Breit-Pauli R-matrix (BPRM) method 
as adopted under the Opacity Project (OP) \cite{op} and the Iron 
Project (IP) \cite{ip} as well as atomic structure calculations
in Breit-Pauli approximation as adopted in program SUPERSTRUCTURE 
(SS) \cite{ss74,ss03} to calculate oscillator strengths and other
transition parameters for photo-excitation and create the absorption
spectra.
We present in the next section a brief outline of the theoretical 
backgrounds that were used to obtain the photo-absorption spectra 
of phosphorus ions.

In the present study, we report photo-absorption features along with
atomic data for bound state energies, radiative transition rates
and lifetimes of all phosphorus ions, P-I through P-XV. For 
accuracy estimation we have compared the present atomic data with those
available in literatures. All atomic data of the present work will be 
available electronically at NORAD-Atomic-Data database \cite{norad}.

\section{Theory}

Discrete lines are formed from atomic photo-excitation. The process of
photo-excitation of ion $X^{+Z}$, where $X$ is the ion of charge $Z$, may be 
expressed as
\begin{equation}
X^{+Z} + h\nu \rightleftharpoons X^{+Z*}
\end{equation}
where $h\nu$ is the photon and $*$ indicates an excited state. The 
process can also lead to a doubly excited state.
The absorption spectral features for all 15 ionization stages of 
phosphorus in the present work were generated using oscillator strengths 
of large number of photo-excitation for each ion. 

Two methodologies are employed here, relativistic Breit-Pauli R-matrix 
(BPRM) method \cite{op,ip} for P~I and P~II, and atomic structure 
calculations using program SUPERSTRUCTURE (SS) in Breit-Pauli 
approximations \cite{ss74,ss03} for all P ions, P~I - P~XV. 
While BPRM can generate a much larger set of dipole allowed E1 
transitions, SS can compute both dipole allowed E1 and forbidden
electric quadrupole (E2), electric octupole (E3), magnetic dipole (M1), 
magnetic quadrupole (M2) transitions. 
Calculations using BPRM codes \cite{betal87,betal95} are {\it ab 
initio} that do not use any model potential, SUPERSTRUCTURE (SS)
\cite{ss74,ss03} implements Thomas-Fermi-Dirac-Amaldi potential 
to represent the electron-electron interactions.

In relativistic Breit-Pauli approximation, the Hamiltonian in Schrodinger
equation,
\begin{equation}
H_{\rm BP}\Psi = E \Psi.
\end{equation}
is given by (e.g. \cite{ss03,aas})
\begin{eqnarray}
 H_{\rm BP}=H^{NR}_{N+1}+H_{\rm mass} + H_{\rm Dar} + H_{\rm so} +  
 \nonumber
 \end{eqnarray}
 \begin{eqnarray}
  \frac{1}{2}\sum_{i\ne j}^N{[g_{ij}(so + so') + g_{ij}(ss')+g_{ij}(css')+
 g_{ij}(d)+g_{ij}(oo')]} 
 \end{eqnarray}
where $H_{N+1}^{\rm NR}$ is the non-relativistic Hamiltonian,
\begin{equation}
H_{N+1}^{\rm NR} = \sum_{i=1}\sp{N+1}\left\{-\nabla_i\sp 2 - 
\frac{2Z}{r_i} + \sum_{j>i}\sp{N+1} \frac{2}{r_{ij}}\right\}, .\label{HNR}
\end{equation}
The next three terms are 1-body correction terms known as the mass correction, 
Darwin and spin-orbit interaction terms,
\begin{eqnarray}
H^{\rm mass} = -{\alpha^2\over 4}\sum_i{p_i^4}\,,\qquad
H^{\rm Dar}={\alpha^2 \over 4}\sum_i{\nabla^2\left({Z\over r_i}\right)}\,,
~~~H_{so} = {Ze^2\hbar^2 \over 2m^2c^2r^3}{\bf L.S}
\end{eqnarray}
and the rest are 2-body terms of spin and orbit interactions. 
$\bf p_i$ is the electron momentum, $\bf L$ and $\bf S$ are orbital 
and spin angular momenta.
Program SUPERSTRUCTURE (SS) \cite{ss74,ss03} includes contributions of 
the three 1-body terms and the 2-body term of the Breit interaction 
(the first three spin-orbit interaction terms), and part of last 
three terms. BPRM codes \cite{betal87,betal95} include the 1-body terms.

Each relativistic correction term improves the accuracy of energy 
levels and hence transitions over the non-relativistic LS coupling
energies and transitions. Hence, SS program improves accuracy by
inclusion of the additional terms in comparison to BPRM method. However, 
accuracy can be enhanced faster with increment in configuration 
interaction than the correction terms. The R-matrix method incorporates
much more configuration interaction than atomic structure calculations. 
Hence, R-matrix results are often more accurate than those of SS.
One main difference in non-relativistic LS 
coupling and relativistic Breit-Pauli approximation is seen as having 
much more fine structure energy levels that belong to a single LS energy 
term. This leads to much more possible transitions among energy levels 
than a single one between two LS terms, and hence creation of a more 
accurate spectrum. The number of transitions is increases even more
since selection rules for fine structure allows more transitions that 
are not possible in LS coupling. In addition, transitions occur among 
the fine structure levels belonging to a LS term which is not possible 
for a LS term. Hence, for accurate spectral features, relativistic
approach is necessary. 

In the BPRM method, the wavefunction expansion is expressed in 
close coupling (CC) approximation, where the atomic system is 
represented as a (N+1) number of electrons, N is the number of 
electrons in the core ion interacting with the (N+1)th electron. 
The total (e+ion) wave function, $\Psi_E$, in a symmetry $SL\pi$ 
is expressed as (e.g.  \cite{aas})
\begin{equation}
\Psi_E(e+ion) = A \sum_{i} \chi_{i}(ion)\theta_{i} + 
\sum_{j} c_{j} \Phi_{j},
\end{equation}
where $\chi_{i}$ is the core ion eigenfunction at the ground and 
various excited levels and the sum is over the number of core ion
excitation considered for the atomic process. $\theta_{i}$ is the 
(N+1)th electron wave 
function with kinetic energy $k_{i}^{2}$ in a channel coupled with 
the core ion labeled as $S_iL_i(J_i)\pi_ik_{i}^{2}\ell_i[SL(J)\pi]$. 
$A$ is the antisymmetrization operator. In the second term which is
basically part of the first term, the $\Phi_j$s are bound channel 
functions of the (N+1)-electrons system that provides the 
orthogonality between the continuum and the bound electron orbitals 
and account for short range correlation.
Substitution of $\Psi_E(e+ion)$ in the Schrodinger equation
introduces a set of coupled equations that are solved by the R-matrix
method. General descriptions of the R-matrix method can be found, e.g.,
in \cite{op,ip,aas}.
The energy eigenvalues from the R-matrix calculations are absolute 
and the (N+1)th electron can be bound or in the continuum depending on 
its negative or positive energy (E).  

In contrast to BPRM, atomic structure calculations, such as using SS, 
compute energy values relative to the ground state, and does not specify 
whether the state is bound or in continuum. The wavefunction in SS is 
similar to the first term of CC expansion, but all core ion orbital 
functions are directly multiplied by the outer electron orbital and the
 sum is  over the configurations producing a  specific state and thus
includes contributions of multi-configurations interactions.
Solutions of from SS calculations are given by Whittaker functions. 
Thomas-Fermi scaling parameters in SS calculations impacts on the 
expanding or compressing the orbital functions but maintains the 
right number of nodes and the orthogonality condition. 

The probability for transition from state $i$ to $j$, $P_{ij}$, due to a
photon absorption is given by (e.g. \cite{aas})
\begin{equation}
P_{ij} = {2\pi}\frac{c^2}{h^2\nu_{ji}^2}|<j |\frac{e}{mc}
{\bf \hat{e}.p}e^{i{\bf k.r}}|i>|^2 \rho(\nu_{ji}).
\end{equation}
where $k$ is the wave vector, $\nu_{ij}$ is the photon frequency for
transition, $\rho$ is the radiation density along with  other standard 
constants.
Various terms in  $e^{i{\bf k.r}}$ introduce various multipole transitions,
such as, the first term gives the electric dipole transitions E1, the
second term gives E2 and M1, and the third term E3 and M2. The general line
strength of the transitions is obtained from
\begin{eqnarray}
S^{{\rm X}\lambda}(ij)&=&
 \Big|\big\langle{\mit\Psi}_j\big\Vert O^{{\rm X}\lambda}\big\Vert
 {\mit\Psi}_i\big\rangle\Big|^2\,,\qquad\quad S(ji)=S(ij).\label{eq:Xlam}
\end{eqnarray}
where $O^{{\rm X}\lambda}$ is the operator for various transitions
${\rm X}\lambda$. We report transitions up to the third term. For E1
transitions, oscillator strengths, radiative decay rate which is also
known as transition probability or Einstein's A-coefficient, and
the corresponding photo-absorption cross sections can be obtained as
\begin{equation}
f_{ij}={E_{ji}\over {3g_i}}S^{\rm E1}(ij), 
\quad A_{ji}^{\rm E1}=\alpha^3{g_i\over g_j}E_{ji}^2f_{ij}\label{eq:Aji},
~\quad \sigma_{PI}(\nu)= 8.064 \frac{E_{ij}} {3g_{i}} S^{\rm E1}~\mbox[Mb], 
 \end{equation}
where $E_{ji}$ is the transition energy, $\nu$ is the photon energy, and
$\alpha$ is the fine structure constant,  $g_j$ and $g_i$ being the
statistical weights of the upper and lower states respectively.
The radiative decay rates for higher order multipole radiation
 electric quadrupole (E2) and magnetic dipole (M1) can be 
obtained as (e.g. \cite{ss03})
\begin{eqnarray}
g_jA^{\rm E2}_{ji}&=&2.6733\times 10^3{\rm s}^{-1}\,(E_j-E_i)^5S^{\rm E2}(i,j)
 \label{eq:E2}\\
g_jA^{\rm M1}_{ji}&=&3.5644\times 10^4{\rm s}^{-1}\,(E_j-E_i)^3S^{\rm M1}(i,j);
 \label{eq:M1}
\end{eqnarray}
and for electric octopole (E3) and magnetic quadrupole (M2) as
\begin{eqnarray}
g_jA^{\rm E3}_{ji}&=&1.2050\times 10^{-3}{\rm s}^{-1}\,(E_j-E_i)^7S^{\rm E3}
(i,j)\label{eq:E3}\\
g_jA^{\rm M2}_{ji}&=&2.3727\times 10^{-2}{\rm s}^{-1}\,(E_j-E_i)^5S^{\rm M2}
(i,j)\,.\label{eq:M2}
\end{eqnarray}
While BPRM can generate a much larger set of transitions for n 
going up to n=10, SS can compute transitions for n going to up to 
5 or 6. 
The accuracy of SS is comparable to that of Dirac-Fock approximation 
for most ions. R-matrix can provide higher accuracy as it can 
accommodate a much larger set of configurations. 

The lifetime of a level $k$ can be computed as
\begin{equation}
\tau_k=\frac{1}{\sum_i A_{ki}}
\end{equation}

%

\section{Computation}

The oscillator strengths for P~I and P~II were obtained using the
Breit Pauli R-matrix package of codes \cite{betal87,betal95}. The
computation involves a number of stages, named as, STG1, STG2, RECUPD, 
STGH, STGB, STGBB. The first stage, STG1, is initiated with 
wavefunctions of the core ion as input and carry out various numerical 
integration. Atomic structure program SUPERSTRUCTURE \cite{ss03} is 
used to obtain the core ion wavefunctions. 

The close coupling wavefunction expansion, Eq.(5), for P~I included  
28 levels of core ion P~II, and that of P~II 18 levels of core ion 
P~III. A set of configurations for each core ion was optimized by
program SS for the wave functions. 
The core ion levels and their calculated energies are presented in 
Table 1.  The table compares the calculated SS energies of the core 
ions with the compiled energies of NIST \cite{nist}. The comparison 
shows agreement between the calculated and measured energies is 
within a few percent for most levels.
 \begin{table}
 \caption{The table presents levels and their energies ($E_t$) of core 
ions P~II  and P~III that are  included in the wave function expansion
 of P~I and P~II, respectively. 
Calculated energies from SUPERSTRUCTURE (SS) are compared with those 
of Martin et al \cite{metal85} available in the NIST table \cite{nist}.}
 \begin{tabular}{rlrlllrll}
 \noalign{\smallskip}
 \hline
 \noalign{\smallskip}
  & \multicolumn{4}{c}{P~II} & \multicolumn{4}{c}{P~III} \\
 & Level & $J_t$ & $E_t$(Ry) & $E_t$(Ry) & Level & $J_t$ & $E_t$(Ry) 
 & $E_t$(Ry)\\
 &  &  &NIST & SS & &  &NIST & SS \\
  \noalign{\smallskip}
  \hline
  \noalign{\smallskip}
 1 & $ 3s^23p^2(^3P  )$   & 0     & 0.0     & 0. &  $ 3s^23p(^2P^o)$   & 1/2   & 0.0        & 0. \\
 2 & $ 3s^23p^2(^3P  )$   & 1   & 0.00150   & 0.00199 & $ 3s^23p(^2P^o)$   & 3/2 & 0.005095  & 0.00429 \\
 3 & $ 3s^23p^2(^3P )$    & 2   & 0.00427   & 0.00562 & $ 3s3p^2(^4P )$    & 5/2 & 0.523559  & 0.51278 \\
 4 & $ 3s^23p^2(^1D )$    & 2   & 0.08094   & 0.10513 & $ 3s3p^2(^4P )$    & 3/2 & 0.520570  & 0.51028 \\
 5 & $ 3s^23p^2(^1S )$    & 0   & 0.19661   & 0.21350 & $ 3s3p^2(^4P )$    & 1/2 & 0.518708  & 0.50874 \\
 6 & $ 3s3p^3(^5S^o)$   & 2   & 0.41642   & 0.34507 &  $ 3s3p^2(^2D )$    & 3/2 & 0.682693  & 0.70551 \\
 7 & $ 3s3p^3(^3D^o)$    & 3  & 0.59461   & 0.59151 &  $ 3s3p^2(^2D )$    & 5/2 & 0.682957  & 0.70565 \\
 8 & $ 3s3p^3(^3D^o)$    & 2  & 0.59481   & 0.59117 &  $ 3s3p^2(^2S )$    & 1/2 & 0.913094  & 0.99414 \\
 9 & $ 3s3p^3(^3D^o)$    & 1  & 0.59512   & 0.59102 &  $ 3s3p^2(^2P )$    & 1/2 & 0.993621  & 1.03335 \\
 10 & $ 3s3p^3(^3P^o)$   & 2  & 0.69953   & 0.70112    & $ 3s3p^2(^2P )$   & 3/2 & 0.997044  & 1.03613     \\
 11 & $ 3s3p^3(^3P^o)$   & 1   & 0.69996   & 0.70162    & $ 3s^23d(^2D )$   & 3/2 & 1.065039  & 1.14381     \\
 12 & $ 3s3p^3(^3P^o)$   & 0   & 0.70006   & 0.70181     & $ 3s^23d(^2D )$   & 5/2 & 1.065142  & 1.14392     \\
 13 & $ 3s3p^3(^1D^o)$   & 2   & 0.70814   & 0.73277     & $ 3s^24s(^2S )$   & 1/2 & 1.073800  & 1.10909     \\
 14 & $ 3s^23p4s(^3P^o )$ & 2   & 0.78913   & 0.80961    & $ 3s^24p(^2P^o )$ & 1/2 & 1.28832   & 1.34867    \\
 15 & $ 3s^23p4s(^3P^o )$ & 1   & 0.79047   & 0.80536    & $ 3s^24p(^2P^o )$ & 3/2 & 1.28956   & 1.34955   \\
 16 & $ 3s^23p4s(^3P^o )$   & 0   & 0.79394   & 0.80394   & $ 3p^3(^2D^o )$   & 3/2 & 1.342508  & 1.38637     \\
 17 & $ 3s^23p3d(^3F^o )$   & 4   & 0.80013   & 0.82044   & $ 3p^3(^2D^o )$   & 5/2 & 1.343073  & 1.38678     \\
 18 & $ 3s^23p3d(^3F^o )$   & 3   & 0.80161   & 0.81777   & $ 3p^3(^4S^o )$   & 3/2 & 1.455433  & 1.49071     \\
 19 & $ 3s^23p3d(^3F^o )$   & 2   & 0.80366  & 0.81585     &  &  &  & \\
 20 & $ 3s^23p4s(^1P^o )$   & 1   & 0.81005 & 0.82108      &  &  &  & \\
 21 & $ 3s^23p4p(^1P   )$   & 1   & 0.92617 & 1.03907      &  &  &  & \\
 22 & $ 3s^23p3d(^1P^o )$   & 1   & 0.93677 & 0.96034      &  &  &  & \\
 23 & $ 3s^23p3d(^3P^o )$   & 2   & 0.94434 & 0.97588      &  &  &  & \\
 24 & $ 3s^23p3d(^3P^o )$   & 1   & 0.94549 & 0.97741      &  &  &  & \\
 25 & $ 3s^23p3d(^3P^o )$   & 0   & 0.94717  & 0.97818     &  &  &  & \\
 26 & $ 3s^23p3d(^3D^o )$   & 1   & 0.94818 & 1.00434      &  &  &  & \\
 27 & $ 3s^23p3d(^3D^o )$   & 3   & 0.94821 & 1.00551      &  &  &  & \\
 28 & $ 3s^23p3d(^3D^o )$   & 2   & 0.94865  & 1.0050      &  &  &  & \\
             \noalign{\smallskip}
 \hline
 \end{tabular}
 \end{table}

As Table 1 shows. for both core ions, outer electron excitation 
included 3d, 4s, 4p orbitals. The final states of the (N+1)-electron
system were obtained from these levels added by the angular 
momenta of the outer electron, ranging between 0$\le l \le$9. Hence
the final states can be $n\le$10. The second term of the wavefunction,
Eq. 6, included 87 configurations for P I and 39 configurations of
P II. 

STG2 and RECUPD carried out the angular algebra, and STGH computed 
the Hamiltonian matrix and dipole transition matrices. 
Computational details for P~II oscillator strengths are similar to 
those for photoionization of the ion described in \cite{snn17}. 

STGB computed the bound states and energies for P~I and P~II. Energy 
eigen values of the Hamiltonian matrix were obtained using fine 
energy meshes to search for the poles. BPRM codes do not identify 
the energy states.  Spectroscopic identifications of the states 
were carried out using an algorithm based on quantum defect theory 
and angular algebra built in a code PRCBPID by Nahar \cite{np00,bpid}.

Program STGBB of the BPRM codes \cite{betal95} was used to compute  
oscillator strengths. The energies and oscillators were processed 
using code PBPRAD \cite{pbprad} and the synthetic spectra were 
produced using code SPECTRA \cite{spec} 

As mentioned above, the transition parameters for other P ions,
including forbidden transitions of P I and P II, were
obtained through atomic structure calculations using the later 
version of the program SUPERSTRUCTURE \cite{ss03}.
For each ion, an optimized set of configurations was selected such
that it would include photo-excitation of low to high temperature 
plasma. SS typically accommodates transitions among configurations 
with principle quantum number $n$ going up to 6. Optimization was
carried out for an overall good agreement of the computed energies with 
those in NIST compilation table. We varied the Thomas-Fermi scaling 
parameters $\lambda_{nl}$ of the orbital wavefunctions for optimization. 
Each configuration was treated spectroscopic, that is, all
configurations were optimized for the best energies..
The focus was on improving energies more in the low to intermediate 
range levels. The final optimized set of configurations and Thomas-Fermi 
$\lambda_{nl}$ parameters are presented in Table 2, 
\begin{table*}
    \caption{Sets of optimized configurations and Thomas-Fermi scaling
parameters ($\lambda_{\it nl}$) for the orbitals used to obtain the
wavefunctions and energies of 15  phosphorus ions using program SS 
\cite{ss03}. In the complete energy tables, configurations are specified 
by their numbers as given within the parenthesis next to them in the table.
}
     \label{tab:table1}
     \begin{tabular} {l l}
                \hline
                \multicolumn{2}{c}{P I}\\
                \hline
  \noalign{\smallskip}
%
  Configurations: &$ 1s^22s^22p^63s^23p^3$(1), $1s^22s^22p^63s^23p^24s$(2),
 $1s^22s^22p^63s3p^4$(3), $1s^22s^22p^63s^23p^24p$(4),\\
 &$1s^22s^22p^63s^23p^23d$(5), $1s^22s^22p^63s^23p^24d$(6), 
 $1s^22s^22p^63s^23p3d^2$(7),\\
 & $1s^22s^22p^63s3p^33d$(8),$1s^22s^22p^63p^5$(9), $1s^22s^22p^63p^43d$(10)\\
 $\lambda_{nl}$&1.40~(1s), 1.25~(2s), 1.20~(2p), 1.27~(3s), 1.02~(3p), 
 1.19~(3d), 1.20~(4s), 0.98~(4p), 1.17~(4d) \\
            \hline
        \multicolumn{2}{c}{P II}\\
            \hline
  \noalign{\smallskip}
 Configurations: &$ 1s^22s^22p^63s^23p^2(1), 1s^22s^22p^63s3p^3(2), 
 1s^22s^22p^63s^23p3d(3), 1s^22s^22p^63s^23p4s(4)$,\\
 & $1s^22s^22p^63s^23p4p(5),1s^22s^22p^63s^23p4d(6)$,$ 1s^22s^22p^63s^23p4f(7), 
 1s^22s^22p^63s^23p5s(8)$,\\
 & $1s^22s^22p^63s3p^23d(9), 1s^22s^22p^63p^4(10), 1s^22s^22p^63s^23d^2(11), 
 1s^22s^22p^63s^23d4s(12),$\\
 &$ 1s^22s^22p^63p^34s(13), 1s^22s^22p^63s3p3d^2(14)$\\ 
 $\lambda_{nl}$&1.42~(1s), 1.25~(2s), 1.18~(2p), 1.15~(3s), 1.1~(3p), 
 1.0~(3d), 1.3~(4s), 0.97~(4p), 1.0~(4d), 1.0~(4f),\\
 & 1.0~(5s)\\
            \hline
            \multicolumn{2}{c}{P III}\\
            \hline
  \noalign{\smallskip}
 Configurations: &$1s^22s^22p^63s^23p(1), 1s^22s^22p^63s3p^2(2), 
 1s^22s^22p^63s^23d(3), 1s^22s^22p^63s^24s(4)$,\\
 &$ 1s^22s^22p^63s^24p(5), 1s^22s^22p^63p^3(6),1s^22s^22p^63s3p3d(7), 
 1s^22s^22p^63s^24d(8)$,\\
 &$1s^22s^22p^63s^24f(9), 1s^22s^22p^63s^25s(10), 1s^22s^22p^63s3p4s(11), 
 1s^22s^22p^63s3p4p(12),$\\
 &$1s^22s^22p^63s3p4d(13), 1s^22s^22p^63s3p4f(14), 1s^22s^22p^63p^23d(15), 
 1s^22s^22p^63p^24s(16)$,\\
 &$1s^22s^22p^63p^24p(17), 1s^22s^22p^63p^24d(18),1s^22s^22p^63p^24f(19)$\\ 
  $\lambda_{nl}$&1.1~(1s), 1.0~(2s), 1.0~(2p), 1.0~(3s), 1.0~(3p), 
  1.0~(3d), 1.1~(4s), 1.0~(4p), 1.0~(4d),\\
  & 1.0~(4f),1.0~(5s), 1.0~(5p)\\
            \hline
            \multicolumn{2}{c}{P IV}\\
            \hline
  \noalign{\smallskip}
  Configurations: &$1s^22s^22p^63s^2(1), 1s^22s^22p^63s3p(2), 
 1s^22s^22p^63s3d(3), 1s^22s^22p^63s4s(4),$\\
 &$1s^22s^22p^63s4p(5)$,$1s^22s^22p^63s4d(6),1s^22s^22p^63s4f(7)$,
 $1s^22s^22p^63p^2(8),$\\
 &$ 1s^22s^22p^63p3d(9), 1s^22s^22p^63p4s(10),1s^22s^22p^63p4p(11),
  1s^22s^22p^63p4d(12)$,\\
 &$1s^22s^22p^63p4f(13),1s^22s^22p^63d^2(14)$,
 $ 1s^22s^22p^63s5s(15), 1s^22s^22p^63s5p(16),$\\
 &$ 1s^22s^22p^63s5d(17), 1s^22s^22p^63s5f(18),$\\
  $\lambda_{nl}$&2.0~(1s), 1.0~(2s), 1.0~(2p), 1.0~(3s), 1.0~(3p), 
 1.0~(3d), 1.1~(4s), 1.0~(4p), 1.0~(4d),\\
 &1.0~(4f),1.0~(5s), 1.0~(5p), 1.0~(5d), 1.0~(5f)\\
            \hline
            \multicolumn{2}{c}{P V}\\
            \hline
  \noalign{\smallskip}
 Configurations: &$1s^22s^22p^63s(1), 1s^22s^22p^63p(2), 1s^22s^22p^63d(3), 
 1s^22s^22p^64s(4), 1s^22s^22p^64p(5)$,\\
 &$ 1s^22s^22p^64d(6), 1s^22s^22p^64f(7),1s^22s^22p^53s^2(8), 
 1s^22s^22p^65s(9), 1s^22s^22p^65p(10),$\\ 
 &$1s^22s^22p^65d(11), 1s^22s^22p^65f(12), 1s^22s^22p^65g(13), 
 1s^22s^22p^43s^23p(14), $\\
 &$1s^22s^22p^43s^23d(15), 1s^22s^22p^53s3p(16), 1s^22s^22p^53s3d(17), 
 1s^22s^22p^53s4s(18),$\\
  &$ 1s^22s^22p^53s4p(19), 1s^22s^22p^53s4d(20),$\\
  $\lambda_{nl}$&1.30~(1s), 1.30~(2s), 0.998~(2p), 1.20~(3s), 0.998~(3p), 
 0.91~(3d), 1.10~(4s), 0.99~(4p),\\
 & 0.995~(4d), 1.0~(4f), 1.0~(5s), 1.0~(5p), 1.0~(5d), 1.0~(5f), 1.0~(5g)\\
        \hline
    \multicolumn{2}{c}{P VI}\\
    \hline
  \noalign{\smallskip}
    Configurations: &$ 1s^22s^22p^6(1), 1s^22s^22p^53s(2), 
    1s^22s^22p^53p, 1s^22s^22p^53d(3), 1s^22s^22p^54s(4),$\\
    &$1s^22s^22p^54p(5), 1s^22s^22p^54d(6),1s^22s^22p^54f(7), 
    1s^22s^22p^55s(8), 1s^22s^22p^55p(9),$\\
    &$1s^22s^22p^55d(10), 1s^22s^22p^55f(11), 1s^22s^22p^55g(12), 
    1s^22s^22p^43s^2(13),$\\
    &$1s^22s^22p^43s3p(14), 1s^22s^22p^43s3p(15), 1s2s^22p^63s(16), 
    1s2s^22p^63p$(17)\\ 
     $\lambda_{nl}$&1.30~(1s), 1.30~(2s), 0.998~(2p), 1.20~(3s), 
     0.998~(3p), 0.91~(3d), 1.10~(4s), 0.99~(4p), \\
     &  0.995~(4d), 1.0~(4f),1.0~(5s), 1.0~(5p), 1.0~(5d), 
     1.0~(5f), 1.0~(5g)\\
    \hline
    \multicolumn{2}{c}{P VII}\\
    \hline
   \noalign{\smallskip}
 Configurations: &$1s^22s^22p^5(1), 1s^22s2p^6(2), 1s^22s^22p^43s(3), 
  1s^22s^22p^43p(4), 1s^22s^22p^43d(5),$\\
  &$ 1s^22s^22p^44s(6), 1s^22s^22p^44p(7), 1s^22s^22p^44d(8)$,
  $1s^22s^22p^44f(9), 1s^22s^22p^45s(10)$,\\
  &$ 1s^22s^22p^45p(11), 1s^22s^22p^45d(12),1s^22s^22p^45f(13), 
  1s^22s^22p^45g(14),1s^22s2p^53s(15)$,\\ 
                \hline
    \end{tabular}
            \end{table*}

            \begin{table*}
     \begin{center}
     \noindent{Table 2 continues.}\\
     \label{tab:table1}
     \begin{tabular} {l l}
  \noalign{\smallskip}
    \hline
  \noalign{\smallskip}
  & $1s^22s2p^53p(16), 1s^22s2p^53d(17), 1s^22s2p^54s(18),
   1s^22s2p^54p(19), 1s^22s2p^54d(20)$,\\
   &$ 1s2s^22p^6(21),1s2s^22p^53s(22), 1s2s^22p^53p(23)$\\ 
     $\lambda_{nl}$&1.30~(1s), 1.20~(2s), 1.15~(2p), 1.10~(3s), 1.10~(3p),
1.10~(3d), 1.0~(4s), 1.0~(4p), 1.0~(4d), \\
 & 1.0~(4f), 1.0~(5s), 1.0~(5p),1.0~(5d), 1.0~(5f), 1.0~(5g)\\
  \noalign{\smallskip}
    \hline
           \multicolumn{2}{c}{P VIII}\\
    \hline
  \noalign{\smallskip}
 Configurations: &$1s^22s^22p^4(1), 1s^22s2p^5(2), 1s^22p^6(3), 
  1s^22s^22p^33s(4),1s^22s^22p^33p(5),$\\
  &$ 1s^22s^22p^33d(6), 1s^22s^22p^34s(7)$,
    $1s^22s^22p^34p(8), 1s^22s^22p^34d(9), 1s^22s^22p^34f(10)$,\\
  &$ 1s^22s^22p^35s(11), 1s^22s^22p^35p(12), 1s^22s^22p^35d(13), 
  1s^22s^22p^35f(14),$\\
    &$1s^22s^22p^35g(15), 1s^22s2p^43s(16), 1s^22s2p^43p(17), 1s^22s2p^43d(18), 
  1s^22s2p^44s(19),$\\
    &$1s^22s2p^44p(20), 1s^22s2p^44d(21),1s^22s^22p^23s^2(22), 1s2s^22p^5(23), 
   1s2s^22p^43s(24),$\\
   &$ 1s2s^22p^43p$(24)\\ 
   $\lambda_{nl}$&1.30~(1s), 1.17~(2s), 1.12~(2p), 0.95~(3s), 1.03~(3p), 
  1.0~(3d), 1.0~(4s), 1.0~(4p), 1.0~(4d),\\
 & 1.0~(4f), 1.0~(5s), 1.0~(5p), 1.0~(5d), 1.0~(5f), 1.0~(5g)\\
    \hline
      \multicolumn{2}{c}{P IX}\\
    \hline
  \noalign{\smallskip}
 Configurations: &$1s^22s^22p^3(1), 1s^22s2p^4(2), 1s^22s^22p^23s(3), 
 1s^22s^22p^23p(4), 1s^22s^22p^23d(5),$\\
    &$1s^22s^22p^24s(6), 1s^22s^22p^24p(7), 1s^22s^22p^24d(8), 
    1s^22s^22p^24f(9), 1s^22s^22p^25s(10),$\\
   &$ 1s^22s^22p^25p(11), 1s^22s^22p^25d(12), 1s^22s^22p^25f(13), 
   1s^22s^22p^25g(14),1s^22p^5,$\\
    &$1s^22s2p^33s(15), 1s^22s2p^33p(16), 1s^22s2p^33d(17), 
    1s2s^22p^4(18),1s2s^22p^33s(19),$\\ 
   &$1s2s^22p^33p(20)$\\ 
     $\lambda_{nl}$&1.35~(1s), 1.25~(2s), 1.15~(2p), 1.20~(3s), 1.15~(3p), 
  1.10~(3d), 1.0~(4s), 1.0~(4p),\\
 &  1.0~(4d), 1.0~(4f), 1.0~(5s), 1.0~(5p), 1.0~(5d), 1.0~(5f), 1.0~(5g)\\
  \noalign{\smallskip}
    \hline
    \multicolumn{2}{c}{P X}\\
    \hline
  \noalign{\smallskip}
    Configurations: &$1s^22s^22p^2(1), 1s^22s2p^3(2), 1s^22p^4(3), 
 1s^22s^22p3s(4), 1s^22s^22p3p(5),$\\
 &$ 1s^22s^22p3d(6), 1s^22s^22p4s(7), 1s^22s^22p4p(8),
    1s^22s^22p4d(9), 1s^22s^22p4f(10),$\\
 &$ 1s^22s^22p5s(11), 1s^22s^22p5p(12), 1s^22s^22p5d(13), 
    1s^22s^22p5f(14), 1s^22s^22p5g(15),$\\
    &$1s^22s2p^23s(16), 1s^22s2p^23p(17), 1s^22s2p^23d(18), 
    1s^22s2p^24s(19), 1s^22s2p^24p(20),$\\
    &$1s^22s2p^24d(21), 1s2s^22p^3(22),1s2s^22p^23s(23), 
    1s2s^22p^23p(24)$\\
     $\lambda_{nl}$&1.42~(1s), 1.25~(2s), 1.15~(2p), 1.25~(3s), 1.15~(3p), 
 1.17~(3d), 1.20~(4s), 1.20~(4p), 1.2~(4d),\\
 &  1.0~(4f),1.0~(5s), 1.0~(5p), 1.0~(5d), 1.0~(5f), 1.0~(5g)\\
    \hline
    \multicolumn{2}{c}{P XI}\\
    \hline
  \noalign{\smallskip}
  Configurations: &$1s^22s^22p(1), 1s^22s2p^2(2), 1s^22p^3(3), 
  1s^22s^23s(4), 1s^22s^23p(5),1s^22s^23d(6),$\\
  &$ 1s^22s2p3s(7), 1s^22s2p3p(8), 1s^22s2p3d(9),1s^2s^24s(10),
    1s^2s^24p(11),$\\
    &$ 1s^2s^24d(12), 1s^2s^24f(13), 1s^2s^25s(14),1s^2s^25p(15),
    1s^2s^25d(16),$\\
   &$ 1s^2s^25f(17), 1s^2s^25g(18), 1s^22s2p4s(19),1s^22s2p4p(20), 
     1s^22s2p4d(21),$\\
    &$ 1s^22s3d^2(22), 1s^22p^23s(23), 
   1s^22p^23p(24), 1s^22p^23d(25), 1s^22p^24s(26), 1s^22p^24p(27),$\\
    &$1s2s^22p^2(28), 1s2s^22p3s(29), 1s2s^22p3p(30), 1s2s^22p3d(31)$\\ 
     $\lambda_{nl}$&2.50~(1s), 1.20~(2s), 1.02~(2p), 1.12~(3s), 1.10~(3p), 
 1.10~(3d), 1.10~(4s), 1.10~(4p), 1.0~(4d), \\
 &  1.0~(4f),1.0~(5s), 1.0~(5p), 1.0~(5d), 1.0~(5f), 1.0~(5g)\\
  \noalign{\smallskip}
    \hline
    \multicolumn{2}{c}{P XII}\\
    \hline
  \noalign{\smallskip}
    Configurations: &$1s^22s^2(1), 1s^22s2p(2), 1s^22s(3), 1s^22s3s(4), 
    1s^22s3p(5), 1s^22s3d(6),1s^22s4s(7),$\\
    &$ 1s^22s4p(8), 1s^22s4d(9), 1s^22s4f(10), 1s^22p3s(11), 
    1s^22p3p(12),1s^22p3d(13),$\\
    &$ 1s^22p4s(14), 1s^22p4p(15), 1s^22p4d(16), 1s^22p4f(17), 
 1s^22s5s(18), 1s^22s5p(19),$\\
    &$1s^22s5d(20), 1s^22s5f(21), 1s^22s5g(22), 1s^22p5s(23), 1s2p5p(24), 
    1s2p5d(25), ,$\\
    &$1s2p5f(26), 1s2p5g(27), 1s2s^23s(28), 1s2s^23p(29), 1s^22p^2(30), 
    1s^23s^2(31), 1s^23p^2(32),$\\
    &$ 1s^23d^2(33)$\\ 
     $\lambda_{nl}$&1.38~(1s), 1.20~(2s), 1.08~(2p), 1.15~(3s), 1.01~(3p), 
     1.0~(3d), 1.0~(4s), 1.0~(4p),
     1.0~(4d), \\
 &  1.0~(4f), 1.0~(5s), 1.0~(5p),1.0~(5d), 1.0~(5f), 1.0~(5g)\\
    \hline
    \multicolumn{2}{c}{P XIII}\\
    \hline
  \noalign{\smallskip}
  Configurations: &$1s^22s(1), 1s^22p(2), 1s^23s(3), 1s^23p(4), 
  1s^23d(5), 1s^24s(6), 1s^24p(7), 1s^24d(8),$\\
    &$1s^24f(9), 1s2s^2(10), 1s2s2p(11), 1s2s3s(12),1s2s3p(13), 1s2s3d(14), 
    1s2s4s(15),$\\
                \hline
    \end{tabular}
     \end{center}
            \end{table*}

            \begin{table*}
     \begin{center}
     \noindent{Table 2 continues.}\\
     \label{tab:table2}
     \begin{tabular} {l l}
  \noalign{\smallskip}
    \hline
  \noalign{\smallskip}
    &$ 1s2s4p(16), 1s2s4d(17), 
    1s2s4f(18), 1s2s5s(19), 1s2s5p(20), 1s2s5d(21), 1s2s5f(22),$\\
    &$1s2s5g(23), 1s2p3s(24), 1s2p3p(25), 1s2p3d(26), 1s2p4s(27), 
    1s2p4p(28), 1s2p^2(29),$\\
    &$ 1s3s^2(30), 1s3p^2(31), 1s3d^2(32)$\\ 
     $\lambda_{nl}$&1.30~(1s), 1.30~(2s), 0.998~(2p), 1.20~(3s), 0.998~(3p), 
 0.91~(3d), 1.10~(4s), 0.99~(4p), 0.995~(4d), \\
 &  1.0~(4f),1.0~(5s), 1.0~(5p), 1.0~(5d), 1.0~(5f), 1.0~(5g)\\
    \hline
    \multicolumn{2}{c}{P XIV}\\
    \hline
  \noalign{\smallskip}
    Configurations: &$1s^2(1), 1s2s(2), 1s2p(3), 1s3s(4), 1s3p(5), 
   1s3d(6), 1s4s(7), 1s4p(8), 1s4d(9), 1s4f(10),$\\
  &$ 2s^2(11), 2p^2(12), 2s^2(13), 3p^2(14), 3d^2(15), 2s2p(16),
   2s3s(17), 2s3p(18), 2s3d(19), $\\
  & $2s4s(20), 2s4p(21),2s4d(22), 2s4f(23), 
  2p3s(24), 2p3p(25), 2p3d(26), 2p4s(27), 2p4p(28)$\\ 
    $\lambda_{nl}$&1.1~(1s), 1.0~(2s), 1.0~(2p), 1.0~(3s), 1.0~(3p), 1.0~(3d), 
   1.0~(4s), 1.0~(4p),\\
    &1.0~(4d), 1.0~(4f), 1.0~(5s), 1.0~(5p)\\
    \hline
    \multicolumn{2}{c}{P XV}\\
    \hline
  \noalign{\smallskip}
    Configurations: &$1s(1), 2s(2), 2p(3), 3s(4), 3p(5), 3d(6), 4s(7), 
   4p(8), 4d(9), 4f(10)$\\ 
    $\lambda_{nl}$&1.0~(1s), 1.0~(2s), 1.0~(2p), 1.0~(3s), 1.0~(3p), 
 1.0~(3d), 1.0~(4s), 1.0~(4p), 1.0~(4d),\\ 
    & 1.0~(4f), 1.0~(5s), 1.0~(5p)\\
    \hline
            \end{tabular}
     \end{center}
            \end{table*}

All SS data were processed using code PRCSS \cite{snn03}. The spectra
of the ion were obtained using the code SPECTRA \cite{spec} that collected
A-values of all dipole allowed transitions and computed the photoabsorption
cross sections. It also added all cross sections corresponding to same
transition wavelengths.
The lifetimes of all levels of each ion were processed using program
LIFETMSS \cite{snn97} which used A-values of all transitions, dipole
allowed and forbidden, to computed the lifetimes.

\section{Results and Discussions}

We present very large sets of atomic data for energies, transition 
parameters (f, S, A-values), lifetimes, and the spectral features in 
various wavelength ranges 
from x-ray to infrared for all 15 ionization stages of phosphorus, P I - 
P XV. In addition to providing accurate atomic data, one primary 
objective of the present study is finding wavelength ranges in which
 prominent spectral features and strong lines exist in P ions and 
hence can be used for search and identification of this bio-signature 
element in the spectra of exoplanets or in any astronomical objects. 
An observed line can be identified if the transitional levels are known. 
However, to predict whether a line can be observed depends on the 
strength of the transition probability, assuming the line is not 
affected by the environmental factors.

Energies and transition parameters for the two ions, P I and P II, have 
been obtained using  the relativistic Breit-Pauli R-matrix (BPRM) method 
while those for ions P~III - P~XV have been obtained in relativistic 
Breit-Pauli approximation using atomic structure code SUPERSTRUCTURE 
(SS). We have computed lifetimes of all excited levels using A-values 
obtained from BPRM method for P~I and P~II, and from SS for P~III-P~XV.
SS has been used for atomic parameters of P~I and P~II as well, 
but we suggest use of BPRM results of P~I and P~II because of the higher
accuracy of the method, and use of SS results only for forbidden transitions 
which the computer package of BPRM method does not compute.

We may mention that BPRM package of codes calculates bound levels with 
$n$ going up to 10. Hence, the spectral lines and features for P I and 
P II correspond to bound-bound transitions only. On the other hand, SS 
computes all possible bound and continuum energy levels arise from 
the given set of configurations.  
Typically the high lying levels from SS exist in the continuum. 
Hence the spectral features includes photoabsorption lines for both 
bound-bound and bound-continuum transitions. The continuum levels, 
beyond the ionization threshold, are the Rydberg autoionizing states, 
and the transitions to continuum appear as resonances in photoionization
cross sections. These lines can be seen observed when they are strong
and isolated. 
Since SS computes the energy of an continuum level as an eigenvalue,  
the resonance is a single point without the typical broadened Lorentzian 
profile.

Accuracy checks of the present atomic parameters for the spectral 
features have been carried out in a number of ways. The atomic data 
have been benchmarked by comparing them with available calculated and 
measured values of energies, transition probabilities, and radiative
lifetimes. 
Radiative lifetime of a level is the reciprocal of sum of transition 
probabilities from the level to its lower levels. It is a measurable 
quantity in laboratories and often measured with high precision. Hence 
lifetimes are commonly used for determination of accuracy of transition 
probabilities. 
Accuracy analysis of the present work indicates that the present spectral 
features can provide dependable guidance for  search of phosphorus.

We compare our energies of P I - P XIV in Table 3 with the measured 
values by Martin et al \cite{metal85} and of P XV with those by Yerokhin 
and Shamaev 
\cite{ys15} and Erickson \cite{erickson77} all of which are listed at 
NIST \cite{nist} table. For brevity, we chose 10 levels of each ion to 
compare, and all comparisons are placed in the same Table 3. The 
table lists the total number of levels obtained for each ion.
Most of 
our energies compared in Table 3 are in excellent agreement, less than 
1 to a few percent, with the measured values. Complete tables of 
energies will be available online at NORAD-Atomic-Data database \cite{norad}.
In the complete table of energies from BPRM method, the configurations 
with the levels are provided. In the energy tables from SS, the configuration 
numbers for the levels are specified. These numbers are the configuration
numbers specified within parenthesis in Table 2. The list of
configurations of Table 2 is also given inside the energy data files.

We discuss the atomic data for transition probabilities, lifetimes and 
spectral features of each ion separately in the subsections below.
However, for brevity with 15 ions of phosphorus, we provide short tables of 
examples of atomic data along with comparison with others for each 
ion. We have combined the short tables for the 15 phosphorus ions 
(P I through P XV) into one table, similar to the energy table, for each 
quantity of transition probabilities and lifetimes. This provides an 
overall picture of the ionization stages of the element.

\subsection{P I}

We have obtained from BPRM method 343 and from SUPERSTRUCTURE 245 
fine structure energy levels of P I, as specified in Table 3. 
We find good agreement in energies, BPRM energies slightly better 
than those from SS, with measured energies of Martin et al 
\cite{metal85}.
The set of 9 configurations for P~I in SS calculations produced 245
levels, including both bound and continuum.
Being neutral P~I was a very sensitive ion for optimization of the 
configurations using SS. 
\begin{table}
  \begin{center}
    \caption {Comparison of the present calculated energies for the
fine structure levels of P-I to P~XV with those of Martin et al 
\cite{metal85} available in the compilation table of the NIST \cite{nist}. 
$N_{BPRM}$, $N_{SS}$ are the total number of energies obtained using 
BPRM approximation and SUPERSTRUCTURE program respectively.
}
   \label{tab:table2}
\begin{tabular} {r l l r l l l}
  \noalign{\smallskip}
                \hline
  \noalign{\smallskip}
\multicolumn{7}{c}{P I:~~\mbox{N$_{BPRM}$=543,~N$_{SS}$=245}}\\
    K & Configuration &Level & J & $E_{NIST}$(Ry) &$E_{SS}$(Ry) &$E_{BPRM}$(Ry)\\
  \noalign{\smallskip}
         \hline
  \noalign{\smallskip}
 1      & $3s^23p^3$    & $^4S^o$& {3/2}        & 0 & 0 & 0. \\
 2      & $3s^23p^3$    & $^2D^o$& {3/2}        & 0.10367& 0.11415 & 0.1108 \\
 3      & $3s^23p^3$    & $^2D^o$& {5/2}        & 0.10353& 0.11417 & 0.1106\\
 4      & $3s^23p^3$    & $^2P^o$& {1/2}        & 0.17084 & 0.16936 & 0.17413\\
 5      & $3s^23p^3$    & $^2P^o$& {3/2}        & 0.17061 & 0.16937 & 0.17429\\
 6      & $3s^23p^24s$  & $^4P^o$& {1/2}        & 0.52352 & 0.48482 &0.53221\\
 7      & $3s^23p^24s$  & $^4P  $& {3/2}        & 0.52369 & 0.48520 &0.53361\\
 8      & $3s^23p^24s$  & $^4P  $& {5/2}        & 0.52393 & 0.52470& 0.53594\\
 9      & $3s^23p^24s$  & $^2P  $& {1/2}        & 0.55908 & 0.54172& 0.5516 \\
10 & $3s^23p^24s$       & $^2P  $& {3/2}        & 0.56001 & 0.54256& 0.55444 \\
                \hline
 \end{tabular}
   \begin{tabular} {r l l r l l l}
%
\multicolumn{7}{c}{P II,~N$_{BPRM}$=475,~N$_{SS}$=243}\\
    K & Configuration &Level & J & $E_{SS}$(Ry) &$E_{NIST}$(Ry) &$E_{BPRM}$(Ry)\\
  \noalign{\smallskip}
         \hline
  \noalign{\smallskip}
   1&$ 3s23p2          $&$^3Pe$& 0.0&    0.000E+00&    0.000E+00&0.000E+00 \\
   2&$ 3s23p2          $&$^3Pe$& 1.0&    1.232E-03&    1.503E-03&1.370E-03 \\
   3&$ 3s23p2          $&$^3Pe$& 2.0&    3.554E-03&    4.275E-03&3.920E-03 \\
   4&$ 3s23p2          $&$^1De$& 2.0&    1.030E-01&    8.094E-02&8.554E-02 \\
   5&$ 3s23p2          $&$^1Se$& 0.0&    2.013E-01&    1.966E-01&2.269E-01 \\
   6&$ 3s3p3           $&$^5So$& 2.0&    3.063E-01&    4.164E-01&4.237E-01 \\
   7&$ 3s3p3           $&$^3Do$& 1.0&    5.835E-01&    5.946E-01&6.190E-01 \\
   8&$ 3s3p3           $&$^3Do$& 2.0&    5.836E-01&    5.948E-01&6.192E-01 \\
   9&$ 3s3p3           $&$^3Do$& 3.0&    5.837E-01&    5.951E-01&6.196E-01 \\
  10&$ 3s3p3           $&$^3Po$& 2.0&    7.602E-01&    6.995E-01&7.476E-01 \\
  \noalign{\smallskip}
                \hline
  \end{tabular}
   \begin{tabular} {r l l r l l }
\multicolumn{6}{c}{P III,~~N$_{SS}$=235}\\
  \noalign{\smallskip}
    K & Configuration &Level & J & $E_{SS}$(Ry) &$E_{NIST}$(Ry) \\
  \noalign{\smallskip}
\hline
  \noalign{\smallskip}
   1&$ 3s23p           $&$^2Po$& 0.5&    0.000E+00&    0.000E+00 \\
   2&$ 3s23p           $&$^2Po$& 1.5&    4.290E-03&    5.095E-03 \\
   3&$ 3s3p2           $&$^4Pe$& 0.5&    5.087E-01&    5.187E-01 \\
   4&$ 3s3p2           $&$^4Pe$& 1.5&    5.102E-01&    5.206E-01 \\
   5&$ 3s3p2           $&$^4Pe$& 2.5&    5.128E-01&    5.236E-01 \\
   6&$ 3s3p2           $&$^2De$& 1.5&    7.056E-01&    6.827E-01 \\
   7&$ 3s3p2           $&$^2De$& 2.5&    7.056E-01&    6.830E-01 \\
   8&$ 3s3p2           $&$^2Se$& 0.5&    9.941E-01&    9.131E-01 \\
   9&$ 3s3p2           $&$^2Pe$& 0.5&    1.0333+00&    9.936E-01 \\
  10&$ 3s3p2           $&$^2Pe$& 1.5&    1.036E+00&    9.970E-01 \\
%
%
  \noalign{\smallskip}
                 \hline
  \noalign{\smallskip}
\multicolumn{6}{c}{P IV, ~~N$_{SS}$=101}\\
  \noalign{\smallskip}
    K & Configuration &Level & J & $E_{SS}$(Ry) &$E_{NIST}$(Ry) \\
  \noalign{\smallskip}
          \hline
  \noalign{\smallskip}
    1&$ 2p63s2          $&$^1Se$& 0.0&    0.000E+00&    0.000E+00 \\
    2&$ 3s3p            $&$^3Po$& 0.0&    6.081E-01&    6.189E-01 \\
    3&$ 3s3p            $&$^3Po$& 1.0&    6.101E-01&    6.210E-01 \\
    4&$ 3s3p            $&$^3Po$& 2.0&    6.140E-01&    6.253E-01 \\
    5&$ 3s3p            $&$^1Po$& 1.0&    9.815E-01&    9.586E-01 \\
    6&$ 3p2             $&$^1De$& 2.0&    1.435E+00&    1.441E+00 \\
\hline
                                        \end{tabular}
                        \end{center}
 \end{table}

  \begin{table}
  \begin{center}
 {Table 3 continues.}\\
    \label{tab:table2}
     \begin{tabular} {r l l r l l}
                 \hline
    K & Configuration &Level & J & $E_{SS}$(Ry) &$E_{NIST}$(Ry) \\
  \noalign{\smallskip}
                 \hline
  \noalign{\smallskip}
    8&$ 3p2             $&$^3Pe$& 1.0&    1.506E+00&    1.505E+00 \\
    9&$ 3p2             $&$^3Pe$& 2.0&    1.510E+00&    1.510E+00 \\
   10&$ 3s3d            $&$^3De$& 1.0&    1.747E+00&    1.726E+00 \\
  \noalign{\smallskip}
                 \hline
  \noalign{\smallskip}
 \multicolumn{6}{c}{P V,  ~~N$_{SS}$=161}\\
  \noalign{\smallskip}
    K & Configuration &Level & J & $E_{SS}$(Ry) &$E_{NIST}$(Ry) \\
  \noalign{\smallskip}
          \hline
  \noalign{\smallskip}
    1&$ 2p63s           $&$^2Se$& 0.5&    0.000E+00&    0.000E+00 \\
    2&$ 2p63p           $&$^2Po$& 0.5&    8.001E-01&    8.077E-01 \\
\hline
                                        \end{tabular}
                        \end{center}
 \end{table}

  \begin{table}
  \begin{center}
 \noindent{Table 3 continues.}\\
    \label{tab:table2}
     \begin{tabular} {r l l r l l}
                 \hline
    K & Configuration &Level & J & $E_{SS}$(Ry) &$E_{NIST}$(Ry) \\
          \hline
  \noalign{\smallskip}
   6&$ 2s2p5           $&$^3Po$& 2.0&    3.753E+00&    3.680E+00 \\
   7&$ 2s2p5           $&$^3Po$& 1.0&    3.811E+00&    3.726E+00 \\
   8&$ 2s2p5           $&$^3Po$& 0.0&    3.841E+00&    3.752E+00 \\
   9&$ 2s2p5           $&$^1Po$& 1.0&    5.302E+00&    5.108E+00 \\
  10&$ 2p6             $&$^1Se$& 0.0&    8.921E+00&    8.598E+00 \\
                \hline
  \noalign{\smallskip}
 \multicolumn{6}{c}{P IX,  ~~N$_{SS}$=498}\\
  \noalign{\smallskip}
    K & Configuration &Level & J & $E_{SS}$(Ry) &$E_{NIST}$(Ry) \\
\hline
                                        \end{tabular}
                        \end{center}
 \end{table}

  \begin{table}
  \begin{center}
 {Table 3 continues.}\\
    \label{tab:table2}
     \begin{tabular} {r l l r l l}
                 \hline
    K & Configuration &Level & J & $E_{SS}$(Ry) &$E_{NIST}$(Ry) \\
          \hline
  \noalign{\smallskip}
 5&$1s22s2p           $&$^1P0$&  1.0&    3.2746E+00& 3.346967\\
   6&$ 1s22p2          $&$^3Pe$& 0.0&    4.410E+00&   4.403E+00 \\
   7&$ 1s22p2          $&$^3Pe$& 1.0&    4.454E+00&    4.440E+00 \\
   8&$ 1s22p2          $&$^3Pe$& 2.0&    4.524E+00&    4.495E+00 \\
   9&$ 1s22p2          $&$^1De$& 2.0&    5.001E+00&    4.936E+00 \\
  10&$ 1s22p2          $&$^1Se$& 0.0&    6.163E+00&    6.057E+00 \\
                \hline
  \noalign{\smallskip}
 \multicolumn{6}{c}{P XIII,  ~~N$_{SS}$=207}\\
\hline
                                        \end{tabular}
                        \end{center}
 \end{table}


We present 32,678 dipole allowed (E1) transitions calculated from 
the BPRM method. A number of transitions along with comparison with 
existing values available in NIST \cite{nist} compiled table are 
presented in Table 4.
For the first dipole allowed transitions, $3s^23p^3(^4S^o) - 3s^23p^24s
(^4P)$, the present A-values from BPRM method have good agreement with 
those of Lawrence \cite{law67} and Fischer et al \cite{fisc2006}. 
However, for the second set of E1 transitions, $3s^23p^3(^4S^o) - 3s3p^4
(^4P)$, there seem to be general disagreement among the groups. For the
rest of the transitions, the present radiative decay rates show
general agreement with Refs\cite{law67,fisc2006} for allowed transitions.

We have computed a smaller set of E1 transitions, a total of 8726, 
using SS. This set includes both the bound-bound and bound-continuum
transitions. Improvement of this set required significant amount of 
optimization because of the sensitivity in slight changes in the 
orbital wavefunctions over the E1 transition parameters. E1 transitions
obtained from SS are also compared in Table 4. Comparison shows general
 agreement. 

We compare the forbidden E2 and M1 transitions obtained from SS with 
those available in the NIST table, calculated by Czyzak and Krueger 
\cite{czy63}. Comparison shows better agreement between the present
and those of Ref.\cite{czy63} for the E2 transitions while present
values for M1 transitions are lower than those of Ref.[\cite{czy63}.
For confirmation of higher accuracy these transitions in P~I need
further study in the future.
We suggest use of BPRM values for E1 transitions which correspond 
to a much larger set of bound levels, and SS values for forbidden 
transitions.  

\begin{table*}
  \begin{center}
    \caption{Comparison of the present A-values in s$^{-1}$ for allowed 
E1 and forbidden E2, M1 transitions in P-I-XV with other published values.
Nt$_{BPRM}$ and Nt$_{SS}$ are the total number of transitions obtained
from BPRM method and SS.}
   \label{tab:table4}
\begin{tabular} {l l r l }
  \noalign{\smallskip}
                \hline
  \noalign{\smallskip}
  & \multicolumn{1}{l}{P I:~~ Nt$_{BPRM}$=32,678,~Nt$_{SS}$=38565}& 
  \multicolumn{2}{c}{$A~(s^{-1})$}\\
  \noalign{\smallskip}
Type & \multicolumn{1}{c}{Transition}  & \multicolumn{1}{c}{Present(BPRM,SS)} & \multicolumn{1}{c}{Others} \\
  \noalign{\smallskip}
         \hline
  \noalign{\smallskip}
E1 & $3s^23p^3(^4S^o_{3/2}) \rightarrow 3s^23p^24s(^4P_{5/2}) $ & 
 2.017E+08,2.22E+08   & $^a$~2.17E+08, $^b$2.043E+08\\
E1 & $3s^23p^3(^4S^o_{3/2}) \rightarrow 3s^23p^24s(^4P^o_{3/2}) $ & 
 2.002E+08,2.26E+08             & $^a$~2.14E+08, $^b$2.013E+08\\
E1 & $3s^23p^3(^4S^o_{3/2}) \rightarrow 3s^23p^24s(^4P^o_{1/2}) $ & 
 1.99E+08,2.28E+08             & $^a$~2.13E+08, $^b$2.013E+08\\
E1  & $ 3s^23p^3(^4S^o_{3/2})\rightarrow 3s3p^4(^4P_{1/2}) $ &  
 1.757E+06,9.27E+07                      & $^c$~3.9E+07, $^b$3.568E+05\\
E1 & $ 3s^23p^3(^4S^o_{3/2})\rightarrow 3s3p^4(^4P_{3/2}) $ &  
 2.105E+06,9.53E+07             & $^c$~4.0E+07, $^b$4.332E+05\\
E1 & $ 3s^23p^3(^4S^o_{3/2})\rightarrow 3s3p^4(^4P_{5/2}) $ &
 2.860E+06,9.99E+07             & $^c$~3.9E+07, $^b$6.229E+05\\
\hline
                                \end{tabular}
    \begin{tabular} {l l r l }
  \noalign{\smallskip}
  &  \multicolumn{1}{l}{P II:~~ Nt$_{BPRM}$=23255,~Nt$_{SS}$=35092}& 
  \multicolumn{2}{c}{$A~(s^{-1})$}\\
  \noalign{\smallskip}
Type& \multicolumn{1}{c}{    Transition} & Present(BPRM,SS) & Others  \\
         \hline
  \noalign{\smallskip}
E1&               $3s^23p^2(^3P_{0})  \rightarrow 3s3p^3(^3D^o_{1})   $ & 
9.023E+06,8.38E+06  & $^a$~7.2E+06, $^g$8.312E+06 \\
E1&                  $3s^23p^2(^3P_{1}) \rightarrow 3s3p^3(^3D^o_{1}) $ & 
6.205E+06,5.40E+06  &$^a$~5.4E+06, $^g$4.905E+06 \\
E1&                   $3s^23p^2(^3P_{1}) \rightarrow 3s3p^3(^3D^o_{2}) $ & 
1.198E+07,1.13E+07  & $^a$~9.6E+06, $^g$1.339E+07 \\
E1&                    $3s^23p^2(^3P_{1}) \rightarrow 3s3p^3(^1D^o_{2})$ & 
2.416E+06,4.38E+05  & $^g$3.249E+06\\
E1&                    $3s^23p^2(^3P_{2}) \rightarrow 3s3p^3(^3D^o_{1}) $ & 
3.529E+05,2.61E+05  & $^a$~3.5E+05, $^g$1.789E+05 \\
E1&                   $3s^23p^2(^3P_{2}) \rightarrow 3s3p^3(^3D^o_{2}) $ &
 3.369E+06,2.76E+06 & $^a$~3.20E+06\\
\hline
\end{tabular}
\end{center}
\end{table*}

\begin{table*}
  \begin{center}
\noindent{Table 4 continues.}\\
   \label{tab:table4}
%
    \begin{tabular} {l l r l }
        \hline
  \noalign{\smallskip}
E2&                      $3s^23p^2(^3P_{1}) \rightarrow 3s^23p^2(^3P_{2})$& 
5.74E-10 &$^d$1.54E-09\\
M1&                      $3s^23p^2(^3P_{1}) \rightarrow 3s^23p^2(^3P_{2})$& 
2.23E-04 &$^e$3.80E-04, $^g$3.15E-04 \\
        \hline
        \multicolumn{4}{l}{$^a:$\cite{savage66}, $^b:$\cite{fuhr98},
                    $^c:$\cite{wiese69}, $^d:$\cite{czy63},
$^g:$\cite{huang85}}\\
  \noalign{\smallskip}
 \hline
  \noalign{\smallskip}
  &  \multicolumn{1}{l}{P III:~~ Nt$_{SS}$=35849}& 
  \multicolumn{2}{c}{$A~(s^{-1})$}\\
  \noalign{\smallskip}
Type& \multicolumn{1}{c}{    Transition} & Present(SS) & Others  \\
         \hline
  \noalign{\smallskip}
E1&                  $ 3s^23p(^2P^o_{1/2})\rightarrow 3s3p^2(^2D_{3/2}) $ & 
6.270E+07 &$^a$~5.5E+07, $^d$5.044E+07\\
E1&                  $3s^23p(^2P^o_{1/2}) \rightarrow 3s3p^2(^2S_{1/2}) $ & 
1.14E+09  &  $^d$8.164E+08\\
E1&                  $ 3s^23p(^2P^o_{1/2})\rightarrow 3s3p^2(^2P_{1/2}) $ & 
4.69E+09  &  $^d$4.474E+09\\
\hline
        \end{tabular}
%
    \begin{tabular} {l l r l }
  \noalign{\smallskip}
  & \multicolumn{1}{l}{P IV:~~ Nt$_{SS}$=6093 }& 
  \multicolumn{2}{c}{$A~(s^{-1})$}\\
  \noalign{\smallskip}
Type& \multicolumn{1}{c}{    Transition} & Present(SS) & Others  \\
         \hline
  \noalign{\smallskip}
E1&                  $3s^2(^1S_{0}) \rightarrow 3s3p(^1P^o_{1}) $ &
4.04E+09  &$^d$~3.94E+09\\
E1&                        $3s3p(^1P^o_{1}) \rightarrow 3p^2(^1D_{2}) $ & 
8.73E+07  &$^d$~1.8E+08, 1.01E+08~$^c$\\
E1&                        $3s3p(^3P^o_{1}) \rightarrow 3p^2(^3P_{2}) $ &
8.69E+08  &$^e$~7.70E+08\\
E1&                        $3s3p(^3P^o_{0}) \rightarrow 3p^2(^3P^o_{1}) $ &
1.15E+09  &$^e$~1.01E+09\\
E1&                        $3p^2(^1D_{2}) \rightarrow 3p3d(^1P^o_{1}) $ &
1.31E+09  &$^d$~4.10E+08\\
E1&                        $3p^2(^3P_{0}) \rightarrow 3p3d(^3D^o_{1}) $ & 
3.64E+09 &$^e$~3.60E+09\\
\hline
  \end{tabular}
%
    \begin{tabular} {l l r l }
  \noalign{\smallskip}
  & \multicolumn{1}{l}{P V:~~ Nt$_{SS}$=17390}& 
  \multicolumn{2}{c}{$A~(s^{-1})$}\\
  \noalign{\smallskip}
Type& \multicolumn{1}{c}{    Transition} & Present(SS) & Others  \\
          \hline
  \noalign{\smallskip}
E1&                 $2p^63s(^2S_{1/2}) \rightarrow 2p^63p(^2P^o_{1/2}) $ & 
1.22E+09  &$^c$~1.16E+09, 1.219E+09~$^b$\\
E1&                $2p^63s(^2S_{1/2}) \rightarrow 2p^63p(^2P^o_{3/2}) $ & 
1.26E+09  &$^c$~1.20E+09, 1.253E+09~$^b$\\ 
E1&                $2p^63p(^2P^o_{1/2}) \rightarrow 2p^63d(^2D_{3/2}) $ & 
3.25E+09 & $^c$~3.10E+09, 4.088E+09~$^a$\\
E1&                       $2p^63p(^2P^o_{1/2}) \rightarrow 2p^64s(^2S_{1/2})$&
2.48E+09   &$^d$~2.5E+09, 2.447E+09~$^b$\\
E1&                     $2p^63p(^2P^o_{3/2}) \rightarrow 2p^64s(^2S_{1/2})$& 
4.98E+09  &$^d$~4.9E+09, 4.913E+09~$^b$\\
E1&                   $2p^63p(^2P^o_{3/2}) \rightarrow 2p^63d(^2D_{5/2}) $ & 
3.84E+09  &$^c$~3.62E+09\\
E1&                    $2p^63p(^2P^o_{3/2}) \rightarrow 2p^63d(^2D_{3/2}) $ &
6.38E+08  &$^c$~6.0E+08\\
E1&                 $2p^63s(^2S_{1/2}) \rightarrow 2p^64p(^2P^o_{3/2}) $ & 
1.53E+09  &$^d$~1.10E+09\\
E1&                 $2p^63s(^2S_{1/2}) \rightarrow 2p^64p(^2P^o_{1/2}) $ & 
1.57E+09  &$^d$~1.10E+09\\
   \hline
 \end{tabular}
 \end{center}
 \end{table*}

 \begin{table*}
   \begin{center}
   \noindent{Table 4. continues.}\\
    \begin{tabular} {l l r l }
   \hline
  \noalign{\smallskip}
E1&                    $2p^63d(^2D_{3/2}) \rightarrow 2p^64f(^2F^o_{5/2})$ & 
8.70E+09  &$^d$~8.80E+09\\
E1&                    $2p^63d(^2D_{5/2}) \rightarrow 2p^64f(^2F^o_{7/2})$ & 
9.32E+09  &$^d$~9.10E+09\\
E1&                    $2p^63d(^2D_{5/2}) \rightarrow 2p^64f(^2F^o_{5/2})$& 
6.21E+08  &$^d$~6.30E+08\\
   \hline
   \multicolumn{4}{l}{$^a$\cite{gode85}, $^b$\cite{john96},
   $^c$\cite{cross65}, $^d$\cite{wiese69}}\\
   \hline
  \noalign{\smallskip}
  & \multicolumn{1}{l}{P VI:~~ Nt$_{SS}$=16731}& 
  \multicolumn{2}{c}{$A~(s^{-1})$}\\
Type& \multicolumn{1}{c}{    Transition} & Present(SS) & Others  \\
  \noalign{\smallskip}
         \hline
  \noalign{\smallskip}
E1&               $2s^22p^6(^1S_{0}) \rightarrow 2s^22p^53s(^1P^o_{1}) $ & 
6.11E+10 &$^a$~4.9E+10, $^b$1.781E+11\\
E1&               $2s^22p^53p(^3S_{1}) \rightarrow 2s^22p^53d(^3P^o_{0})$& 
3.004E+09 & $^b$2.761E+09\\
\end{tabular}
  \noalign{\smallskip}
  & \multicolumn{1}{l}{P VII:~~ Nt$_{SS}$=93962}& 
  \multicolumn{2}{c}{$A~(s^{-1})$}\\
Type& \multicolumn{1}{c}{    Transition} & Present(SS) & Others  \\
  \noalign{\smallskip}
         \hline
  \noalign{\smallskip}
E1&                 $2s^22p^5(^2P^o_{3/2}) \rightarrow 2s2p^6(^2S_{1/2}) $& 
2.658E+10 &$^a$~2.9E+10, $^c$2.557E+10 \\
E1&                $2s^22p^5(^2P^o_{3/2}) \rightarrow 2s^22p^43s(^2P_{3/2})$&
 1.133E+11 & $^c$1.114E+11\\
E1&                $2s^22p^5(^2P^o_{3/2}) \rightarrow 2s^22p^43s(^4P_{5/2})$&
 3.069E+08 &$^c$1.797E+08\\
E1&                $2s^22p^5(^2P^o_{3/2}) \rightarrow 2s^22p^43s(^4P_{3/2})$&
 4.775E+09 &$^c$2.840E+09\\
E1&             $2s^22p^5(^2P^o_{3/2}) \rightarrow 2s^22p^43s(^4P_{1/2})$& 
1.956E+07 &$^c$7.583E+06\\
E1&                $2s^22p^5(^2P^o_{3/2}) \rightarrow 2s^22p^43s(^2P_{1/2})$& 
4.900E+10 & $^c$4.762E+10\\
\hline
    \end{tabular}%
  \end{center}
\end{table*}


Lifetimes of a number of levels of P~I 
were measured using beam-foil technique by Curtis et al \cite{curtis71}
and using fluorescence detection technique by Berzinish et al 
\cite{berz1997}. Present calculated lifetimes computed using A-values 
from BPRM method are compared with measured values in Table 5. We find
excellent agreement with the measured values and other theoretical 
values by Froese-Fischer et al \cite{fisc2006}.
%
\begin{table*}
  \begin{center}
    \caption {Comparison of present calculated radiative lifetimes 
$\tau$ (in 10$^{-9}$ s) of levels of P ions with available 
experimental values using beam-foil (BF) and fluorescence detection (FD) 
techniques, storage ring (SR), Cyclotron (CR)  and other theoretical values.
Complete tables of lifetimes of all excited levels will be available online.}
       \label{tab:table4}
   \begin{tabular} {r l c c c c c}
                \hline
  \noalign{\smallskip}
\multicolumn{7}{c}{P I}\\
  \noalign{\smallskip}
 & & &  & \multicolumn{3}{c}{$\tau$ (ns)}\\
   {K} & \multicolumn{1}{c}{Configuration} & \multicolumn{1}{c}{Level} &
Present (BPRM) &Experiment & Method & Theory\\
         \hline
  \noalign{\smallskip}
           1    & $3s^23p^24s$  & $^2D_{5/2}$ & 3.45 &$ 3.6\pm0.4~^a$&BF & $3.58~^c$\\
        2       & $3s^23p^24s$  & $^4P_{5/2}$ & 4.956&$ 4.8\pm0.5~^a$&BF & $4.89~^c$\\
      3         & $3s^23p^24p$  & $^4S^o_{3/2}$&33.11&$ 36.9\pm1.8~^b$ & FD & $27.75~^c$\\
     4          & $3s^23p^24p$  & $^2D^o_{3/2}$&51.11 &$41.6 \pm14.8~^b$ & FD & $50.2~^c$\\
      5         & $3s^23p^24p$  & $^4P^o_{3/2}$&39.46 &$ 38.9\pm6.2~^b$&FD & $34.9~^c$\\
      6         & $3s^23p^24p$  & $^4D^o_{3/2}$& 48.19, &$ 48.1\pm8.2~^b$&FD & $44.03~^c$\\
      7         & $3s^23p^24p$  & $^4D^o_{5/2}$& 47.89 &$ 48.4\pm5.5~^b$&FD & $43.84~^c$\\
      8         & $3s^23p^24p$  & $^4D^o_{7/2}$& 45.47 &$ 48.3\pm4.4~^b$ &FD & $43.80~^c$\\
  \noalign{\smallskip}
      \hline
       \multicolumn{7}{l}{$^a$\cite{curtis71},$^b$, \cite{berz1997},$^c$\cite{fisc2006}}\\
        \hline
  \noalign{\smallskip}
        \multicolumn{7}{c}{P II}\\
  \noalign{\smallskip}
         \hline
  \noalign{\smallskip}
   1 & $3s3p^3$      & $^3D^o_{3}$ & 58.96   &$ 45\pm10~^b$     &BF&$93.8~^c$\\
   2   & $3s3p^3$      & $^3P^o_{1}$   &21.37 &$ 14\pm 0.8~^a$&BF &$12.22~^ c$\\
   3   & $3s3p^3$      & $^3P^o_{2}$   & $ 25.63 $&$ 14.6\pm0.5~^a$ &BF&$12.98~^c$\\
  &     &                 &             & $9.0\pm0.5~^b$  &BF&         \\
\hline
\multicolumn{7}{l}{$^a$\cite{brown18}, $^b$\cite{curtis71}, $^c$\cite{fisc2006}}\\
                \hline
\multicolumn{7}{c}{P III}\\
  \noalign{\smallskip}
         \hline
  \noalign{\smallskip}
1       & $3s3p^2$      & $^2P_{3/2}$   &$ 0.13 $&$ 0.20\pm0.05~^a$ &BF &$0.15~^b$\\
      2         & $3s3p^2$      & $^2D_{5/2}$   & $ 13.16 $&$ 14\pm2.0~^a$ &BF &$15.2~^b$\\
      3         & $3p^3$        & $^2D_{5/2}$   &$ 7.49 $&$ 1.8\pm0.4~^a$&BF &$11.18~^b$\\
      4         & $3s^24p$      & $^2P^o_{3/2}$         &$ 2.14 $&$ 2.8\pm0.2~^a$&BF &$2.41~^b$\\
      5         & $3s^24f$      & $^2F^o_{7/2}$         &$ 0.91 $&$ 0.70\pm0.15~^a$&BF &$$\\
      6         & $3s3p4s$      & $^4P^o_{5/2}$         &$ 0.65 $&$ 0.5\pm0.2~^a$ &BF &$$\\
\hline
\multicolumn{7}{l}{ $^a$\cite{curtis71}, $^b$\cite{fisc2006}}\\
                \hline
\multicolumn{7}{c}{P IV}\\
         \hline
  \noalign{\smallskip}
       1        & $2p^63s3p$    & $^1P^o_{1}$   &$ 0.241 $&$ 0.22\pm0.02~^a $&BF &$0.27~^d$\\
       2 & $2p^63p^2$   & $^3P_{2}$     &$ 0.289 $&$ 0.32\pm0.03~^a$&BF &$0.32~^d$\\
                3 & $2p^63p^2$  & $^1D_{2}$     &$ 10.8 $&$ 8.2\pm0.8~^a$&BF &$9.88~^d$\\
           4  & $2p^63s3d$      & $^3D_{3}$     &$ 0.193 $&$ 0.36\pm0.05~^a $&BF &$0.21~^d$\\
      5         & $2p^63s4p$    & $^3P^o_{2}$   &$ 1.12 $&$ 1.23\pm0.09~^c$ &BF &$1.15~^d$\\
      6         & $2p^63s4p$    & $^3P^o_{1}$   &$ 1.08 $&$ 1.22\pm0.09~^c $&BF &$1.14~^d$\\
      7         & $2p^63s4d$    & $^3D_{1}$     &$ 1.66 $&$ 1.75\pm0.2~^c $&BF &$1.44~^d$\\
          8 & $2p^63s4f$        & $^3F^o_{4}$   &$ 0.176 $&$ 0.4\pm0.1~^a $&BF &$0.17~^d$\\
     9  & $2p^63s5p$    & $^3P^o_{2}$   &$ 1.15 $&$ 2.5\pm0.15~^c $&BF &$$\\
     10         & $2p^63s5p$    & $^3P^o_{1}$   &$ 1.26 $&$ 2.48\pm0.15~^c$&BF &$$\\
         11     & $2p^63p3d$    & $^1D_{2}$     &$ 0.239 $&$ 0.5\pm0.3 ~^a$&BF &$0.27~^d$\\
         12     & $2p^63p4s$    & $^3P^o_{0}$   &$ 0.207 $&$ 0.27\pm0.04~^b $&BF &\\
\hline
\multicolumn{7}{l}{$^a$ \cite{curtis71}, $^b$ \cite{khayat19},  
                   $^c$ \cite{west89}, $^d$ \cite{fisc2006}}\\
                \hline
\multicolumn{7}{c}{P V }\\
         \hline
  \noalign{\smallskip}
   1       & $2p^63p$      & $^2P^o_{3/2}$         &$ 0.794 $&$ 0.70\pm0.15~^a $&BF &$0.78~^c$\\
  2       & $2p^63d$      & $^2D_{5/2}$   &$ 0.261 $&$ 0.38\pm0.06~^b $&BF &$0.26~^c$\\
 3  & $2p^64s$      & $^2S_{1/2}$   &$ 0.134 $&$ 0.32\pm0.05~^b$&BF &$0.13~^c$\\
\hline
                \end{tabular}
                  \end{center}
                    \end{table*}

\begin{table*}
  \begin{center}
 \noindent{Table 5 continues.}\\
  \begin{tabular} {c c c c c c c}
                \hline
\multicolumn{7}{c}{P V }\\
 & & &  & \multicolumn{3}{c}{$\tau$ (ns)}\\
   {K} & \multicolumn{1}{c}{Configuration} & \multicolumn{1}{c}{Level} &
Present &Experiment & Method & Theory\\
         \hline
  \noalign{\smallskip}
    4   & $2p^64f$      & $^2F^o_{7/2}$         &$ 0.107 $&$ 0.17\pm0.03~^b $&BF &$0.11~^c$\\
    5   & $2p^65f$      & $^2F^o_{7/2}$         &$ 0.198 $&$ 0.38\pm0.06~^b$&BF &$0.19~^c$\\
    6   & $2p^65g$      & $^2G_{9/2}$   &$ 0.375 $&$ 1.56\pm0.23~^b $&BF &$$\\
  \noalign{\smallskip}
\hline
  \noalign{\smallskip}
\multicolumn{7}{l}{ $^a$\cite{curtis71}, $^b$\cite{maio76},
$^c$\cite{fisc2006}}\\
                \hline
  \noalign{\smallskip}
\multicolumn{7}{c}{P VI}\\
  \noalign{\smallskip}
                \hline
  \noalign{\smallskip}
                1       & $1s^22s^22p^53s$      & $^1P^o_{1}$   &$ 0.0164 $&$ 0.018\pm0.002~^a $&BF &$0.0169~^b$\\
      2         & $1s^22s^22p^53s$      & $^3P^o_{1}$   &$ 0.0704 $&$ 0.105\pm0.007~^a $&BF &$0.106~^b$\\
  \noalign{\smallskip}
\hline
  \noalign{\smallskip}
\multicolumn{7}{l}{$^a$ \cite{trabert96}, $^b$ \cite{hibb93}}\\
                \hline
  \noalign{\smallskip}
\multicolumn{7}{c}{P VIII}\\
  \noalign{\smallskip}
                \hline
  \noalign{\smallskip}
                1       & $1s^22s^22p^4$        & $^1D_{2}$     & 1.70E07 &$ 2.86E07\pm8.0E04~^a $&SR & $2.45E07~^b$\\
  \noalign{\smallskip}
         \hline
\multicolumn{7}{l}{$^a$ \cite{trabert12}, $^b$ \cite{cheng79}}\\
         \hline
  \noalign{\smallskip}
\multicolumn{7}{c}{P IX}\\
  \noalign{\smallskip}
                \hline
  \noalign{\smallskip}
                1       & $1s^22s^22p^3$        & $^2P^o_{1/2}$ &$ 7.01E06 $&$ 1.01E07\pm1.2E05~^a $&BF &$9.95E06~^b$\\
        2       & $1s^22s^22p^3$        & $^2P^o_{3/2}$         & 2.90E06 &$ 4.2E06\pm1.0E05~^a $&SR & $4.22E06~^b$\\
  \noalign{\smallskip}
         \hline
\multicolumn{7}{l}{ $^a$ \cite{trabert12}, $^b$ \cite{cheng79}}\\
  \noalign{\smallskip}
                \hline
  \noalign{\smallskip}
\multicolumn{7}{c}{P X}\\
  \noalign{\smallskip}
                \hline
  \noalign{\smallskip}
                1       & $1s^22s^22p^2$ & $^1D_{2}$    &$ 1.21E+07 $&$ 1.77E07\pm3.5E04~^a $&SR &$1.549E07~^c$\\
      2         & $1s^22s2p^3$  & $^3S^o_{1}$     &$ 0.022    $&$ 0.028\pm0.004~^b $&BF &$0.023~^d$\\
      3         & $1s^22s2p^3$  & $^3P^o_{1}$   &$ 0.132  $&$ 0.17\pm0.01~^b $&F &$0.172~^d$\\
      4         & $1s^22s2p^3$  & $^3P^o_{2}$   &$ 0.136  $&$ 0.6\pm0.1~^b$&BF &$0.176~^e$\\
      5         & $1s^22s2p^3$  & $^3D^o_{1}$     &$ 0.330  $&$ 0.42\pm0.02~^b$&BF &$0.323~^d$\\
      6         & $1s^22p^4$    & $^1D_{2}$     &$ 0.0714   $&$ 0.082\pm0.006~^b$&BF &$0.07~^f$\\
      7         & $1s^22p^4$    & $^3P_{2}$     &$ 0.0367  $&$ 0.045\pm0.005~^b$&BF &$0.038~^f$\\
  \noalign{\smallskip}
         \hline
\multicolumn{7}{l}{$^a$ \cite{trabert12}, $^b$ \cite{trabert80a},
$^c$ \cite{cheng79}, $^d$ \cite{wiese69}, $^e$ \cite{nico73},
$^f$\cite{faw79}}\\
                \hline
%
  \noalign{\smallskip}
\multicolumn{7}{c}{P XI}\\
  \noalign{\smallskip}
                \hline
  \noalign{\smallskip}
        1               & $1s^22s2p^2$  & $^2P_{1/2}$   &$0.04854 $&$ 0.055\pm0.006~^a $&BF &$0.052~^b$\\
        2               & $1s^22s2p^2$  & $^2D_{3/2}$   &$0.3596 $&$ 0.11\pm0.03~^a $&F &$0.392~^c$\\
               &                          &                        &              &$ 0.425\pm0.03~^a $&                                 \\
                3               & $1s^22p^3$& $^2P^o_{3/2}$&$ 0.05982 $&$ 0.073\pm0.04~^a $&BF &$0.074~^d$\\
        4               & $1s^22p^3$            & $^4S^o_{3/2}$ &$ 0.0513 $&$ 0.062\pm0.06~^a $&BF &$0.059~^c$\\
              &                          &                        &              &$ 0.32\pm0.05~^a $     &                                 \\
  \noalign{\smallskip}
         \hline       
\multicolumn{7}{l}{ $^a$ \cite{trabert80a}, $^b$ \cite{wiese69},
                   $^c$ \cite{flower75}, $^d$ \cite{dank76}}\\
                \hline
\multicolumn{7}{c}{P XII}\\
  \noalign{\smallskip}
                \hline
  \noalign{\smallskip}
                1       & $1s^22s2p$    & $^1P^o_{1}$   &$ 0.131 $&$ 0.14\pm0.01~^a $&BF &$0.12~^b$\\
      2         & $1s^22p^2$    & $^1S_{0}$     &$ 0.0869 $&$ 0.089\pm0.01~^a $&BF &$0.095~^b$\\
      3         & $1s^22p^2$    & $^3P_{2}$     &$ 0.17 $&$ 0.18\pm0.01~^a $&BF &$0.169~^b$\\
      4         & $1s^22p^2$    & $^1D_{2}$     &$ 0.7598 $&$ 0.79\pm0.05~^a $&BF &$0.75~^b$\\
  \noalign{\smallskip}
         \hline
  \noalign{\smallskip}
\multicolumn{7}{l}{$^a$ \cite{trabert80b}, $^b$ \cite{faw78}}\\
                \hline
                \end{tabular}
                  \end{center}
                    \end{table*}

\begin{table*}
  \begin{center}
 \noindent{Table 5 continues.}\\
  \begin{tabular} {c c c c c c c}
                \hline
\multicolumn{7}{c}{P XIII }\\
 & & &  & \multicolumn{3}{c}{$\tau$ (ns)}\\
   {K} & \multicolumn{1}{c}{Configuration} & \multicolumn{1}{c}{Level} &
Present &Experiment & Method & Theory\\
         \hline
  \noalign{\smallskip}
  1  & $1s2s2p$ & $^4P^o_{5/2}$  &$ 0.1622 $&$ 1.2\pm0.1~^a $&CR &$1.3~^b$\\
  \noalign{\smallskip}
         \hline
\multicolumn{7}{l}{$^a$ \cite{desc82}, $^b$ \cite{cheng74}}\\
  \noalign{\smallskip}
                \hline
 \noalign{\smallskip}
\multicolumn{7}{c}{P XIV}\\
  \noalign{\smallskip}
                \hline
 \noalign{\smallskip}
   1    & $1s2p$     & $^3P^o_{0}$  &$ 6.90 $& $ 4.8\pm0.4~^a$ &BF &\\
   2    & $1s2p$     & $^3P^o_{2}$  &$ 2.87 $&$3.6\pm0.1~^b$&CT &$3.5~^c$\\
 \noalign{\smallskip}
         \hline
\multicolumn{7}{l}{ $^a$\cite{living82}, $^b$\cite{desc82},
                   $^c$\cite{fisc2006}}\\
 \noalign{\smallskip}
\hline
\end{tabular}
        \end{center}
        \end{table*}


Features of photoabsorption spectrum of P~I is presented in Fig. 1.
Panel (a) shows the spectrum going up to 1$\times 10^6~\AA$
wavelengths covering most of the strong lines of P~I. The figure 
shows the wavelength regions of strengths, particularly up to 
up to far infra-red region of 4.5$\times 10^5~\AA$ that dominate the 
spectrum. It is somewhat surprising that the strengths of lines 
continue to remain strong at 1M $\AA$, although density of strong 
lines has become sparse. The upper panel (b) elaborates the spectrum up to
31,000 $\AA$. a range for James Webb Space Telescope (JWST), giving
wavelength regions of strong line strength. This includes optical and infra-red
regions. However, the strengths are lower that those in the far-infrared
regions. 
\begin{figure*}
   \begin{center}
 \includegraphics[width=6.00in,height=3.750in]{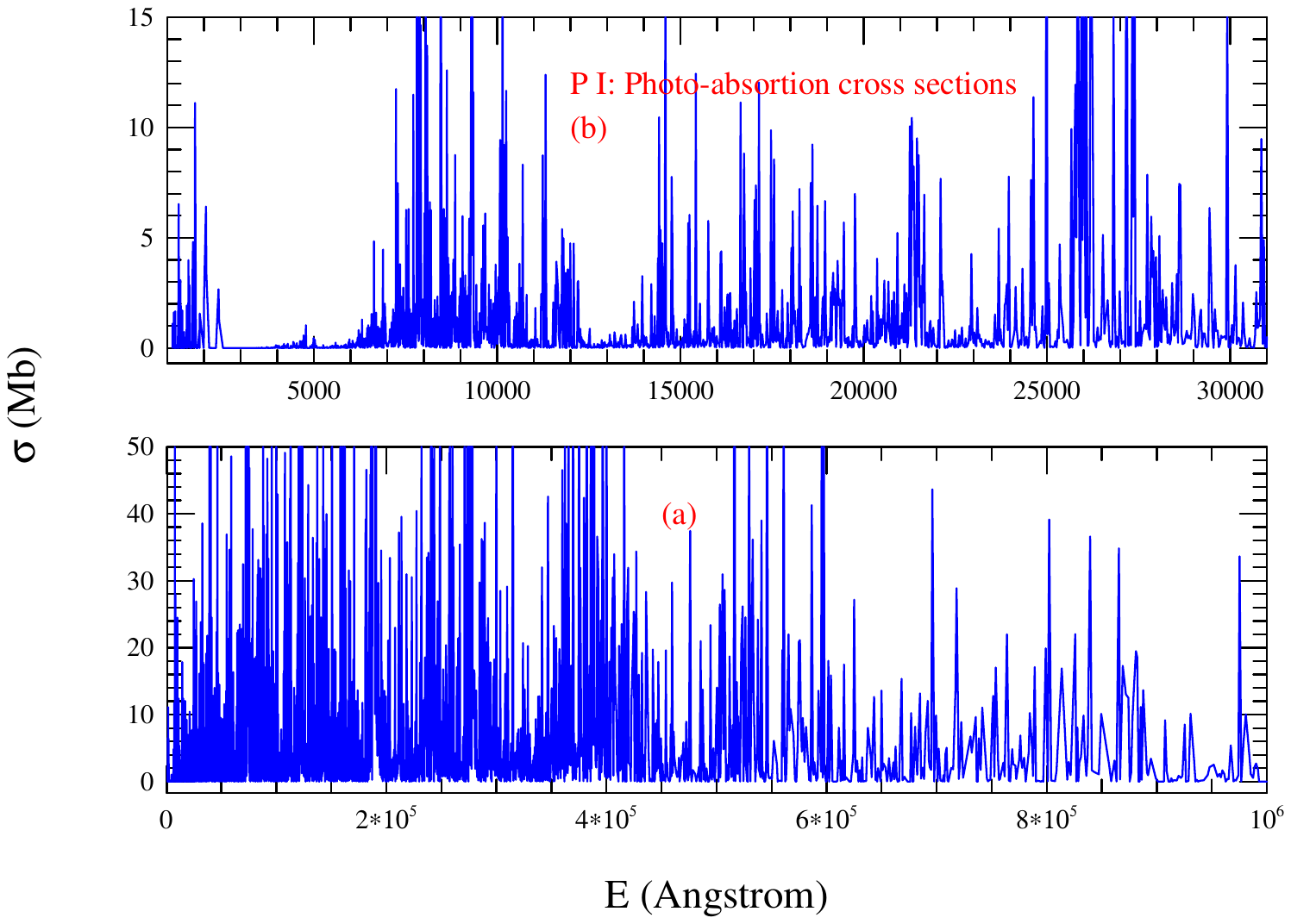}
\caption{\label{fig: p1}
Complete spectrum of P~I for strong lines using 32,678 E1 transitions. 
The lower panel (a)
presents line strength continuing up to far infra-red region of 
1M $\AA$ and reveals dominance in the optical and infra-red regions. The 
upper panel (b) elaborates the line features up to 31000 $\AA$. This 
region shows wavelength regions of dominance. However, the intensity 
is lower than that in the far IR region.
}
\end{center}
\end{figure*}

\subsection{P II}

We have obtained for P~II, 475 bound fine structure levels from BPRM 
method and 243 bound and continuum levels from SUPERSTRUCTURE. 
The BPRM energy levels are from the 18 levels of the core ion 
combining with the outer electron of angular momenta l=0-9 and $n\le$10.
A small set of energy levels from both approaches are listed in Table 3
and compared with those of measured by Martin et al \cite{metal85} 
and available in NIST \cite{nist}. BPRM values are slightly higher and
SS values are slightly lower than the measured values. The overall
agreement of them with the measured values is in general is good.
With inclusion of more configuration interactions, as explained
for the case of P~I, the wavefunctions
for BPRM method are expected to be more accurate. 

We have obtained a larger set of E1 transitions, 23255, from BPRM
method compared to 7920 from SS. and 27172 forbidden transitions from SS.
Limited number of transitions were studied by a number of investigators.
Comparison is made with them in Table 4.
With similar reasons given for P~I, we suggest use of BPRM transitions
for P~II which are expected to be more accurate. A-values for the 
forbidden transitions should be used from SS calculations.

We compare the present A-values from BPRM method with the existing ones.
Although general agreement, particularly with orders of magnitude, are
found among various calculations, differences are also noted. One 
reason for discrepancy was noted from identification of levels. The
mixed levels often have different leading percentages from different
methods toward a LS state and can cause unrealistic discrepancy by 
the different identifications. This appears to be one main reason for
differences in agreement for P~II A-values. 
Although A-values of E1 transitions 
from BPRM method are preferable, we compare A-values of from SS also in
in the table tor accuracy estimation of their A-values for the 
forbidden transitions. We compare the present A-values from SS for 
E2 and M1 transitions with available values, and the agreement is in
general good, except for $3s^23p^2(^3P_1)- 3s^23p^2(^3P_2)(E2)$  
given that these transitions are much more weaker than E1 transitions.  

We have found lifetimes of three levels belonging to configuration 
$3s3p^3$ of P~II, measured by two groups \cite{curtis71,brown18} and 
theoretically computed values by one group \cite{fisc2006}. We 
compare the present values with them in Table 5.  The present lifetime 
for $^3D^o_3$, 59 ns is close to the upper limit of experimental value 
of 55 ns \cite{curtis71}. However, 94 ns predicted by Froese-Fischer 
et al \cite{fisc2006} is much higher than the present as well as the
experimental value. The other two lifetimes show a of values where the 
present lifetimes for $^3P^o_{1,2}$, about 21 and 25 ns, are higher 
than those, about 14 ns by Curtis et al \cite{curtis71} which are again 
higher than the measured value of about 9 ns by Brown et al \cite{brown18} 
and theoretical prediction of about 12 ns by \cite{fisc2006}. Some more 
study of lifetimes may confirm the accuracy of the values.

Features of photoabsorption spectrum of P~II are presented in Fig. 2.
Panel (a) shows the spectrum going up to 5$\times 10^5~\AA$,
wavelengths covering most of the strong lines of P~II. The figure
shows that the wavelength regions of line strengths, particularly up to
far infra-red region of 3$\times 10^5~\AA$ that dominate the
spectrum and beyond which the density of strong lines is reduced. 
The upper panel (b) elaborates the spectrum up to
35,000 $\AA$. a range covering for JWST. We note the density and 
strengths of lines dominate the region of 1700 to 35000 $\AA$.
This includes optical and infra-red regions. However, similar to P~I,
the strengths are lower than those in the far-infrared regions.
\begin{figure*}
   \begin{center}
 \includegraphics[width=6.00in,height=3.75in]{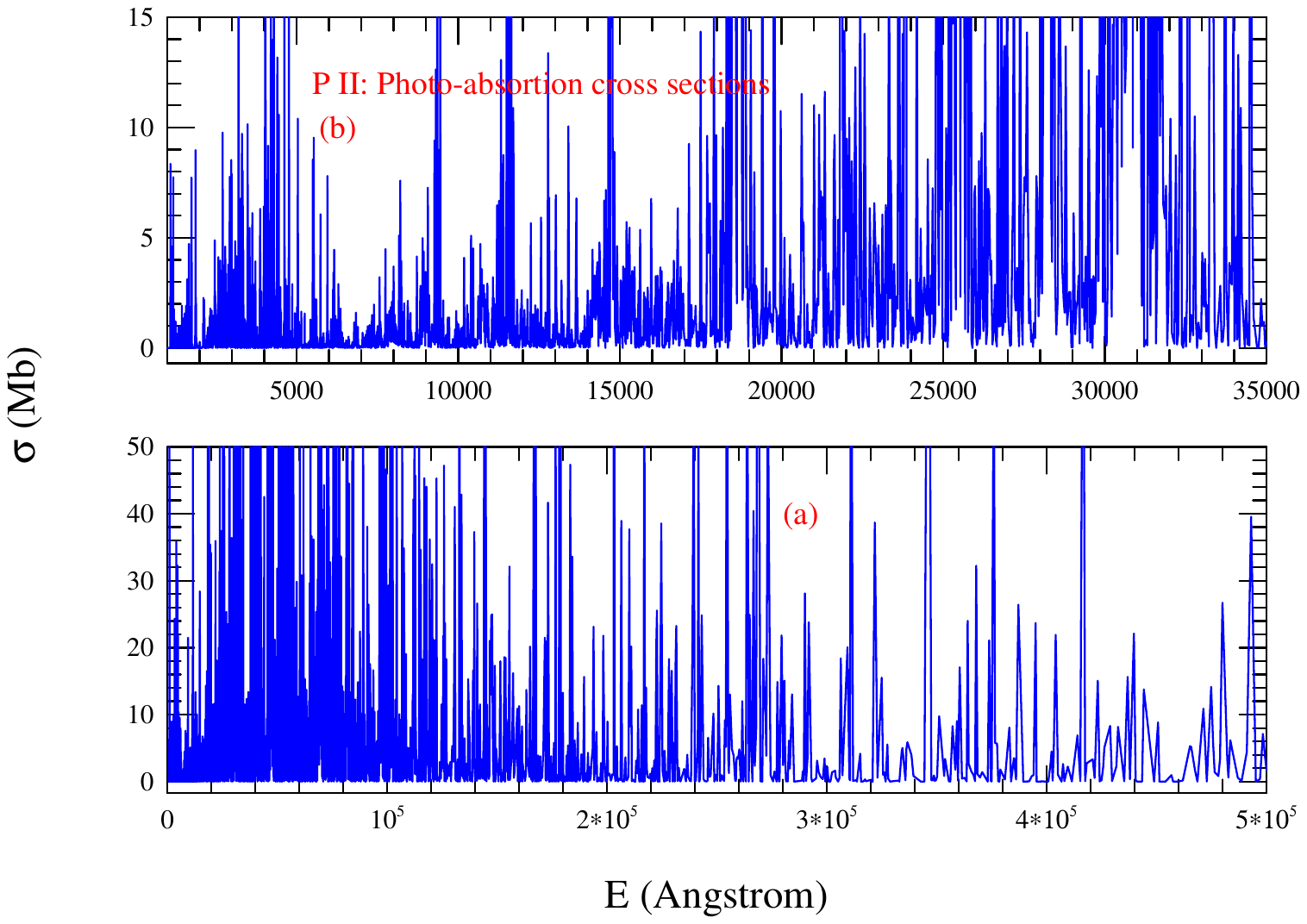}
\caption{\label{fig: p1}
The complete spectrum of P~II for strong lines plotted with 23255
E1 transition. The lower panel (a)
presents strengths up to far infra-red region of 0.5M $\AA$ which
indicate dominant regions up to 280000 $\AA$. The upper panel (b)
elaborates dominant regions in optical and IR wavelengths.
}
\end{center}
\end{figure*}

\subsection{P III}

For the given set of configurations in Table 2, we have obtained 235
fine structure levels. Comparison in Table 3 shows differences of
the calculated energies with the measured values are mainly in the
third figure.

We report a total of 35849 transition rates for Al-like phosphorus 
(P III) and 8441 of them are of type E1 transitions. A-values 
are compared with existing calculated values in Table 4. 
The present values comparable with Huang \cite{huang86}, Wiese et al 
\cite{wiese69} and Fuhr et al \cite{fuhr98}. For the forbidden transition,
$3s^23p(^2P^o_{1/2}) - 3s^23p(^2P^o_{3/2})$, the present A-value 9.38E-04  
$s^{-1}$ is lower and that of Huang is 5.688E-03 $s^{-1}$ which is higher 
than 1.57E-03 $s^{-1}$ of Naqvi \cite{naqvi51} which is rated as A in the 
NIST table.

For P III, lifetimes from 6 excited levels have been presented; two 
levels from the $3s3p^2$ configuration and one level from each of the 
configurations: $3p^3$, $3s^24p$, $3s^24f$ and $3s3p4s$. The present 
values compare very well with both experimental and the other theoretical 
values except for one even level $^2D_{5/2}$. The present value
is higher than the experimental one. Predicted lifetime of Frose-Fischer 
is even higher than the present value. 

Photoabsorption spectrum of P~III is presented in Fig. 3. Unlike 
P I and P II, the spectrum does not extend to infra-red (IR) region. All
strong lines appear in different regions of x-ray to  ultraviolet
wavelengths range. 
 The reason for missing lines in the IR region could be not including 
transitions beyond than 5s. 
\begin{figure*}
   \begin{center}
 \includegraphics[width=6.00in,height=2.50in]{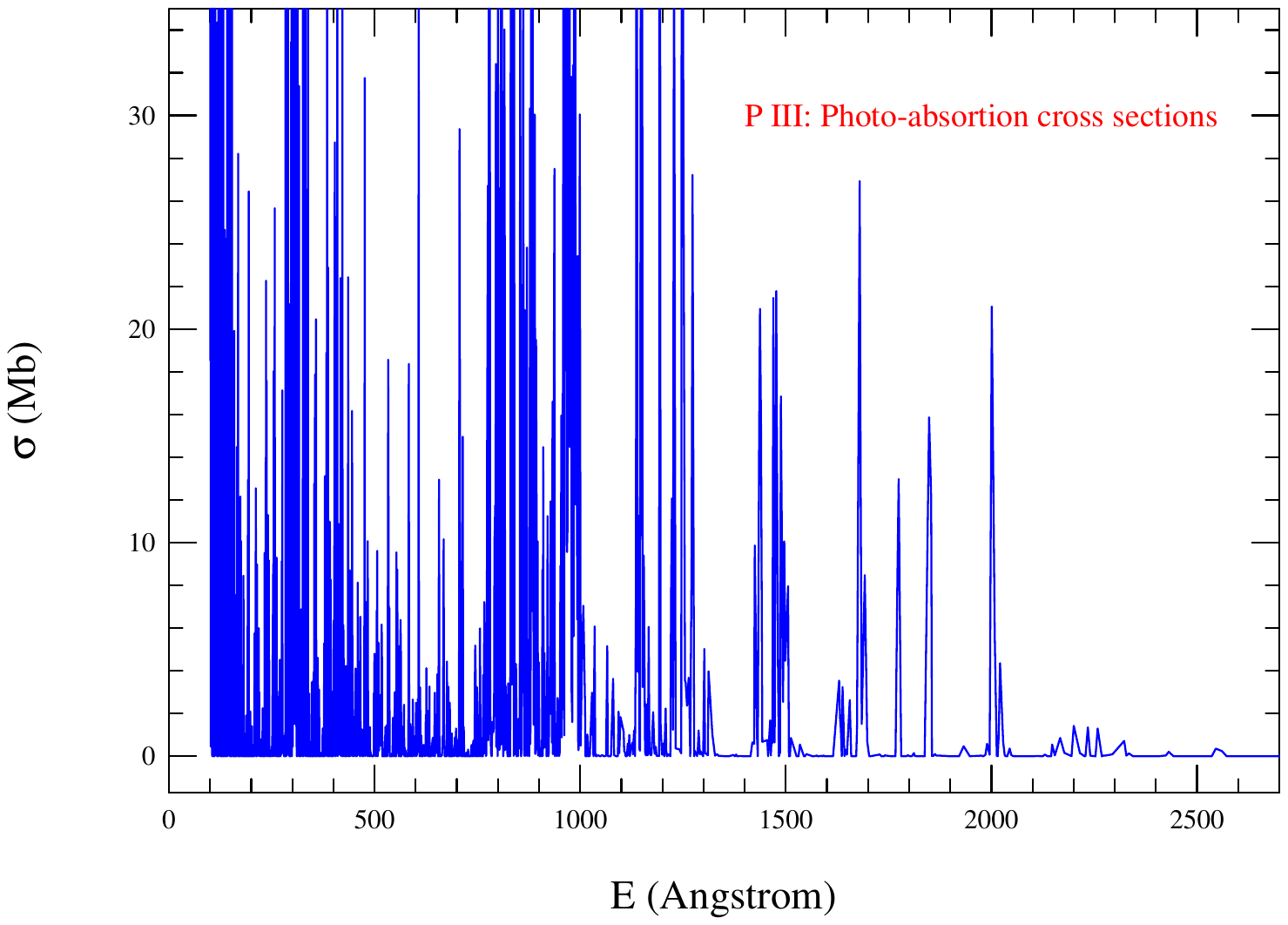}
\caption{\label{fig: p3}
The photoabsorption spectrum of P~III plotted with 8441 E1 transitions
demonstrates presence of strong lines in the x-ray to ultra-violet 
wavelength regions.
}
\end{center}
\end{figure*}

\subsection{P IV}

From 18 configurations with orbitals going up to 5f, we have obtained
101 levels of P IV from SS. They agree very well with the measured 
values of Martin etl al \cite{metal85}, presented in Table 3.

We have obtained a total of 6093 allowed and forbidden transition out 
of which 1501 are E1 transitions. We compare the transition probabilities  
with those calculated by others \cite{zare67,cross65,gode85,naqvi51} 
and find very good agreement with them. The A-values of weak M1 transitions 
from the present work are also in very good agreement except one, 
$3s3p(3Po_1) - 3s3p(1Po_1)$, with those of Naqvi \cite{naqvi51}.

For P IV, lifetimes from 12 excited levels have been presented in
Table 5. Our values compare very well with both experimental values of
\cite{curtis71,khayat19,west89} except for one odd level $^4P$ and 
two even levels $^4S$ and $^2D$.
However they agree with all theoretical values by Frose-Fischer
et al. \cite{fisc2006}.

\noindent
Photoabsorption spectrum:

Photoabsorption spectrum of P~IV  is presented in Fig. 4. As seen in
the spectrum, spectral lines dominate the x-ray to ultraviolet wavelengths
regions. However, P~IV shows some dominant lines in the optical region
going to near IR region.
\begin{figure*}
   \begin{center}
 \includegraphics[width=6.00in,height=2.75in]{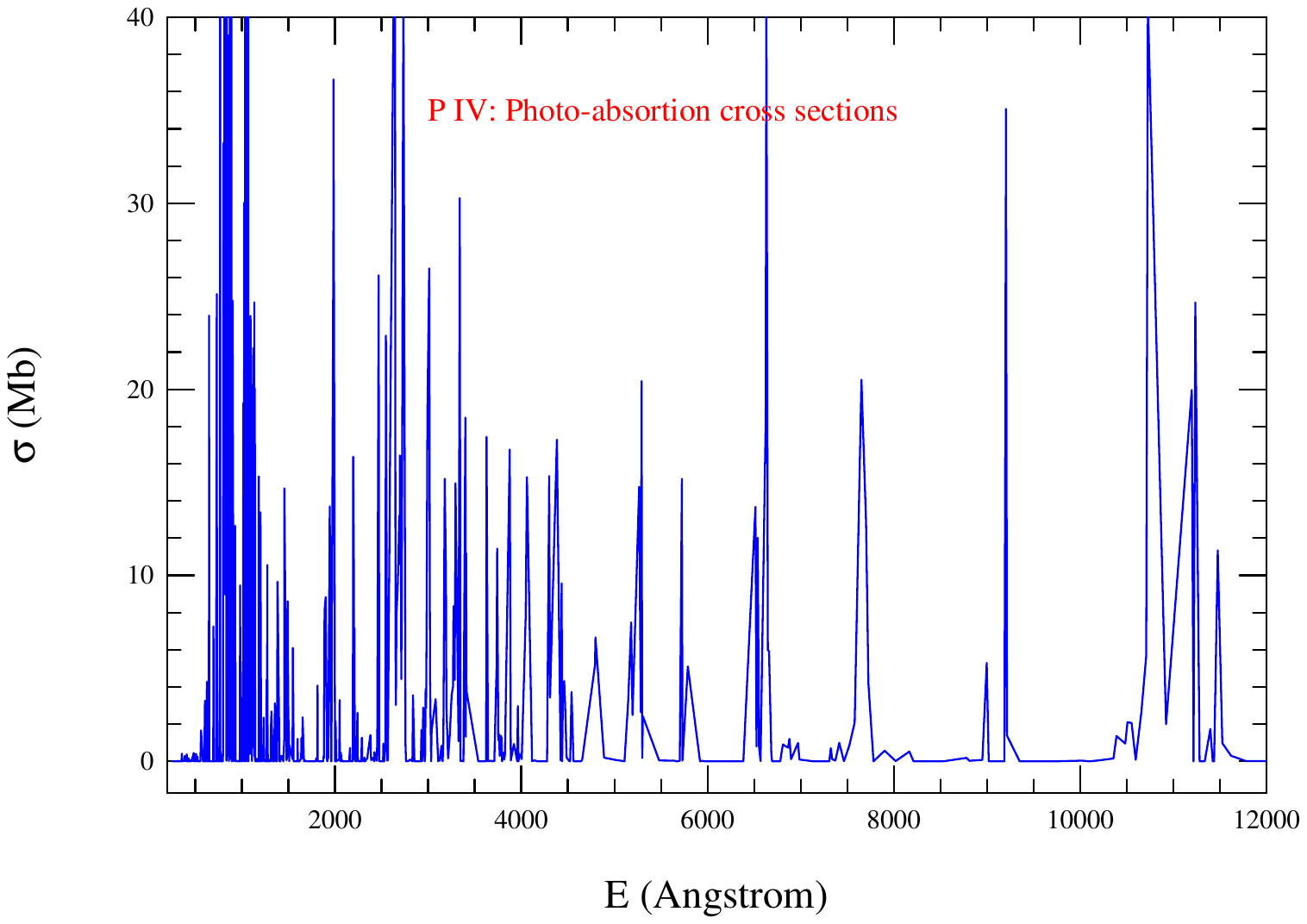}
\caption{\label{fig: p3}
The spectrum of P~IV  exhibiting  strong lines in the x-ray to 
ultra-violet wavelength regions, and some isolated strong lines in the
visible to near IR regions. A total of 1501 E1 transitions have been
included in the spectrum.
}
\end{center}
\end{figure*}

\subsection{P V}

We have obtained 161 fine structure levels of P~V from the given set
of configurations going up to 5g given in Table 2. The energies have 
excellent agreement with the measured values of Martin et al 
\cite{metal85} in Table 3.

Transition rates for Na-like phosphorus (P V) have been calculated 
for 17390 transitions of which 4309 are E1 transitions.
The values are compared in Table 4 with those calculated by
others \cite{gode85,john96,cross65,wiese69} with very good agreement. 

For P V, lifetimes from 6 excited levels have been presented in Table
5.  Our values compare very well with those predicted values of 
Frose-Fischer et al \cite{fisc2006}. Agreement is good with 
experimental values of \cite{curtis71},\cite{maio76} except of 
a couple. 

Photoabsorption spectrum of P~V  is presented in Fig. 5. The 
spectrum indicates dominant spectral lines in regions of soft x-ray 
to far ultraviolet wavelengths regions. The spectrum also indicates 
presence of some dominant lines in the optical region.
\begin{figure*}
   \begin{center}
 \includegraphics[width=6.00in,height=2.75in]{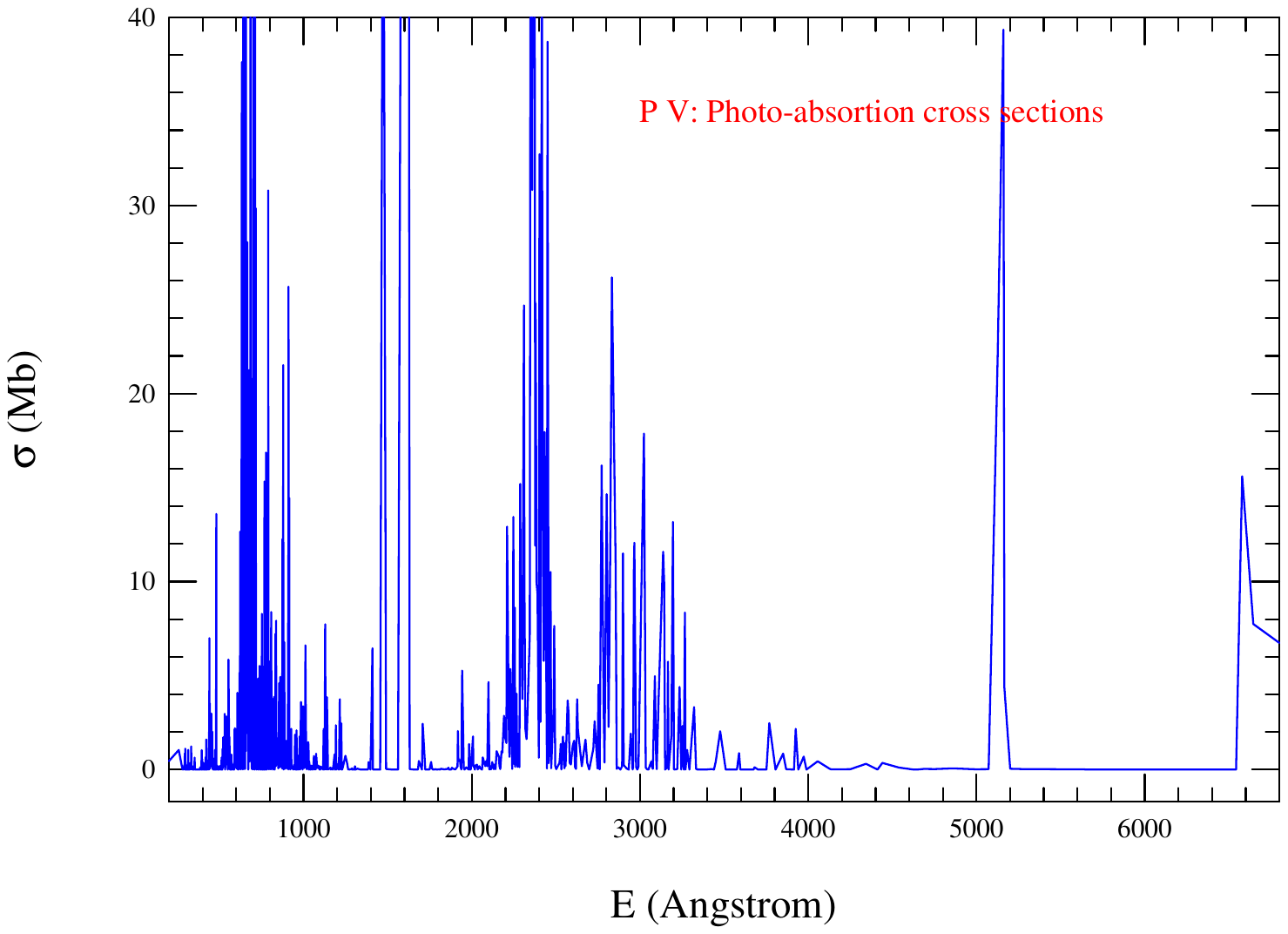}
\caption{\label{fig: p5}
The spectrum of P~V  exhibiting  strong lines in the regions of x-ray 
to far ultra-violet wavelength regions, and some isolated strong lines 
in the visible region. A total of 4309 E1 transition have been included
in the spectrum. 
}
\end{center}
\end{figure*}

\subsection{P VI}

We have obtained 168 fine structure levels of P~VI which are in less
than 1\% with most of the measured values of Martin et al \cite{metal85}.

Transition rates for Ne-like phosphorus (P VI) have been calculated for 
16731 transition of which 3687 are E1 transitions.
Examples are presented with comparison with those by Kaster et al 
\cite{kastner67} (NIST included their values in the compiled table) and
by Hibbert et al \cite{hibb93} in Table 4. The present values compared 
much better with with those of Kastner et al than of Hibbert et al
\cite{hibb93}. 

For P VI, we compare the present lifetimes with lifetimes of 
2 excited levels $3s^23p^23s(^1P^o_1,^3P^o_1)$ measured by Trabert
\cite{trabert96} and calculated by Hibbert et al \cite{hibb93} in Table
5. Our value for $^1P^o_1$ compares very well with both experimental
and theoretical ones. For the other level, our value is slightly lower
than the measured range and that of Hibbert et al. 

Photoabsorption spectrum of P~VI is presented in Fig. 6. The 
regions of strong spectral lines extends from x-ray to infra-red (lower
panel). The upper panel elaborates the region up to optical wavelengths.
\begin{figure*}
   \begin{center}
 \includegraphics[width=6.00in,height=2.75in]{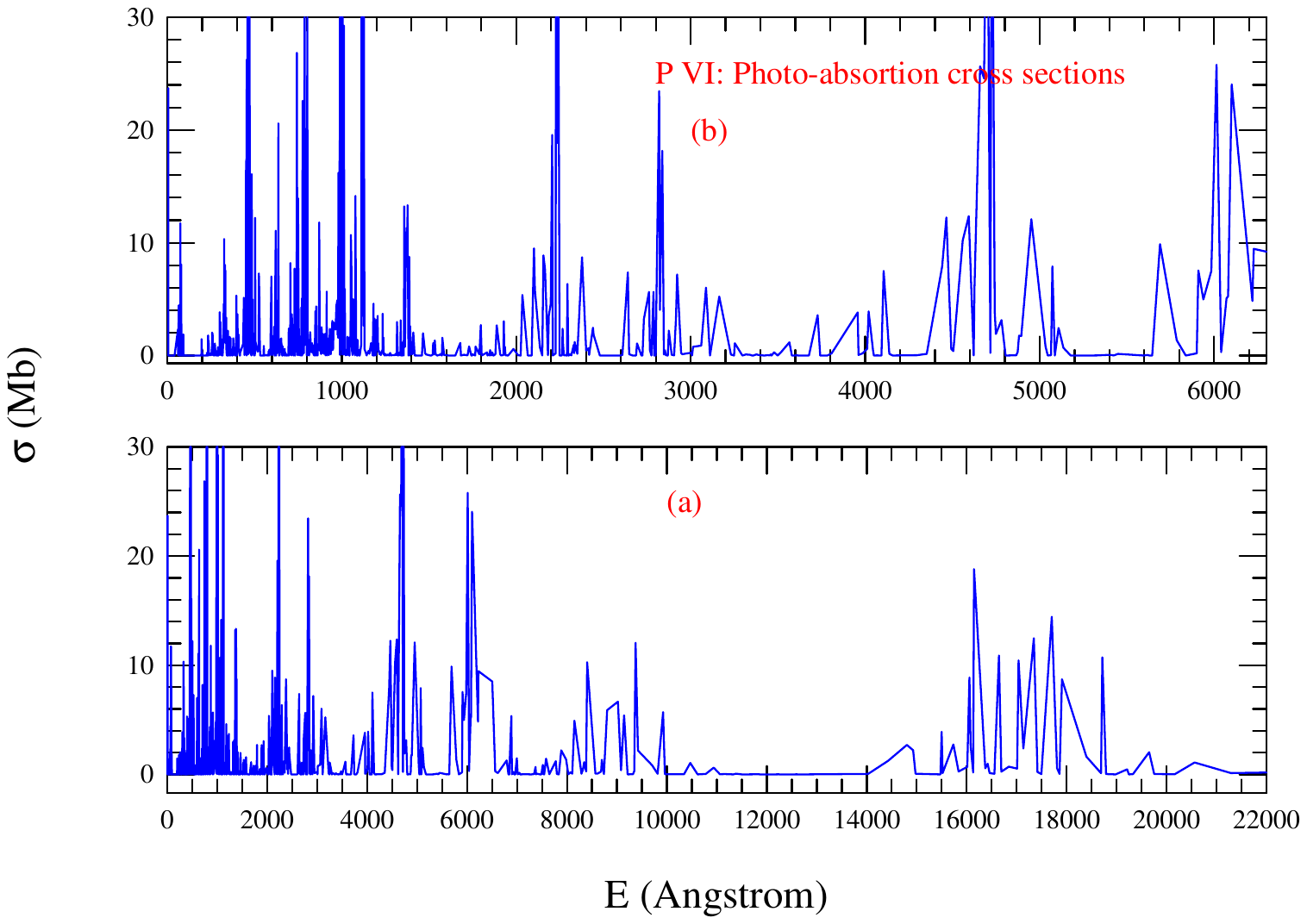}
\caption{\label{fig: p6}
The spectrum of P~VI, created with 3687 E1 transitions is exhibiting  
strong lines in the regions of x-ray to infra-red wavelengths in
panel A. The upper panel elaborates the region of x-ray to optical. 
}
\end{center}
\end{figure*}

\subsection{P VII}

We have obtained 386 fine structure levels for P~VII which show large
differences with the measured values of Martin et al \cite{metal85} 
as presented in Table 3.

Transition rates for F-like phosphorus (P VII) have been calculated 
for 93962 transitions with 21897 are of type E1. The values are 
compared with those calculated by Aggarwal \cite{agg2019}i, Cohen and
Dalgarno \cite{cohen64} with very good agreement in Table 4. However,
a single forbidden M1 transition by Naqvi \cite{naqvi51} listed in the 
NIST has much lower value of 6.89 s$^{-1}$ compared to the present
value of 15 s$^{-1}$.

We have a large set of lifetimes for P~VII, and have not found any 
published value to compare with.

Photoabsorption spectrum of P~VII is presented in Fig. 7. The 
regions of strong spectral lines extends from x-ray to infra-red (lower
panel). The upper panel elaborates the region up to optical wavelengths.
\begin{figure*}
   \begin{center}
 \includegraphics[width=6.00in,height=3.50in]{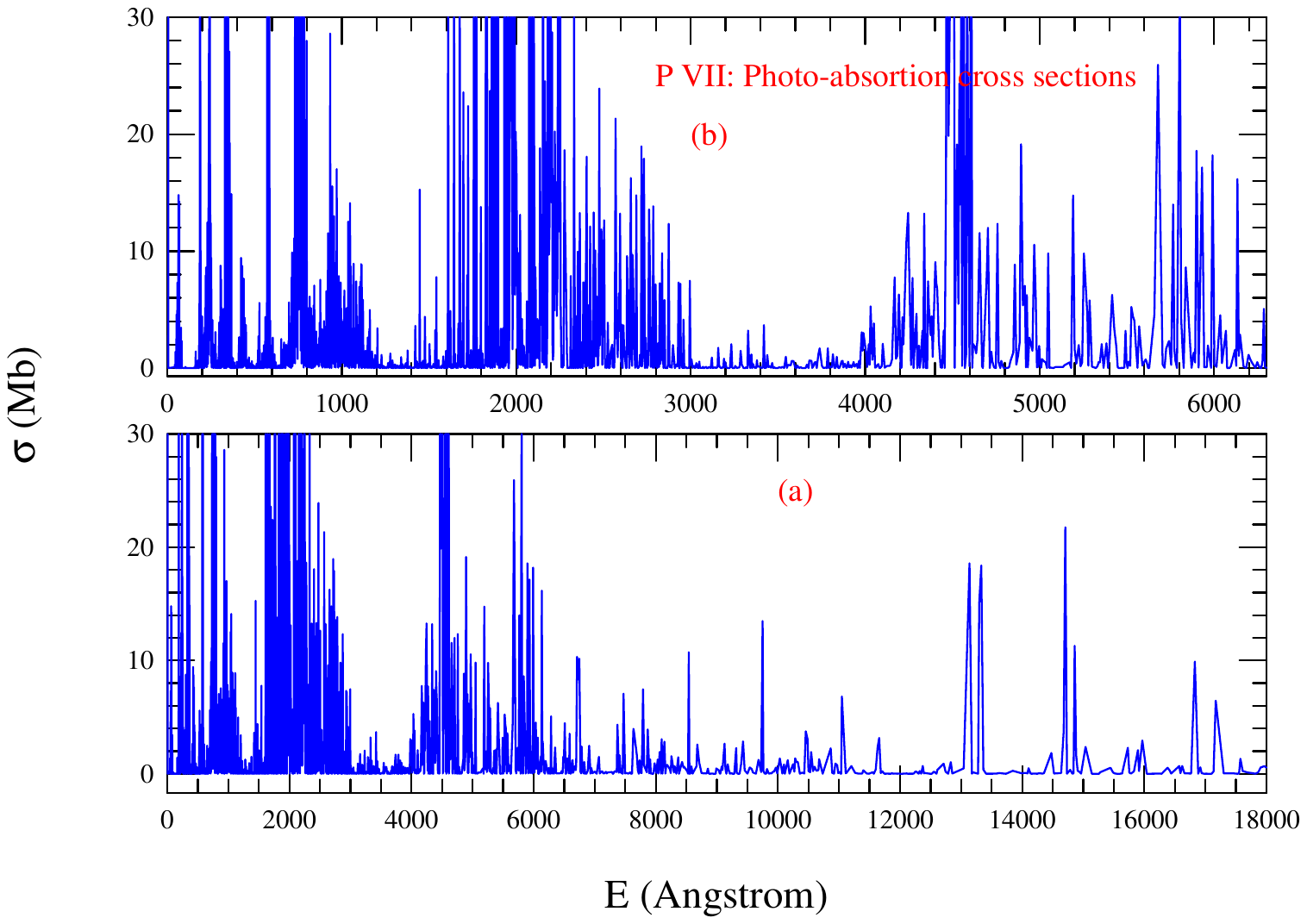}
\caption{\label{fig: p7}
The spectrum of P~VII, created with 21897 E1 transitions, is exhibiting  
strong lines in the regions of x-ray to infra-red wavelengths in panel 
(a). The upper panel (b) elaborates the region of x-ray to optical. 
}
\end{center}
\end{figure*}

\subsection{P VIII}

We have obtained 653 fine structure levels of P~VIII which are in less
than 1 to a few percent differences with the measured
values of Martin et al \cite{metal85} presented in Table 3.

Transition rates for O-like phosphorus (P VIII) have been calculated for 
120624 transitions with 59219 are of type E1.
The values are compared in Table 4 with those calculated by
Cheng \cite{cheng79}, Cohen and Dalgarno \cite{cohen64}. We find very 
good agreement between the present and reported values for the allowed
transitions. For forbidden transitions, there are some differences
with Naqvi \cite{naqvi51} and Malville and Berger \cite{malv65} but 
overall in agreement is acceptable as these transitions are very weak,

For P VIII, lifetime for 1 excited level from the ground configuration,
$3s^23P^4(^1D_2)$ are presented. The lifetime is calculated from the
forbidden transition. Our value 1.70E07 ns is comparable but lower than  
the measured lifetime 2:86E07 ns \cite{trabert12} and calculated value
by Cheng \cite{cheng79}.

Photoabsorption spectrum of P~VIII is presented in Fig. 8. The 
regions of strong spectral lines extends from x-ray to infra-red (lower
panel). The upper panel elaborates the region up to optical wavelengths.
\begin{figure*}
   \begin{center}
 \includegraphics[width=6.00in,height=3.50in]{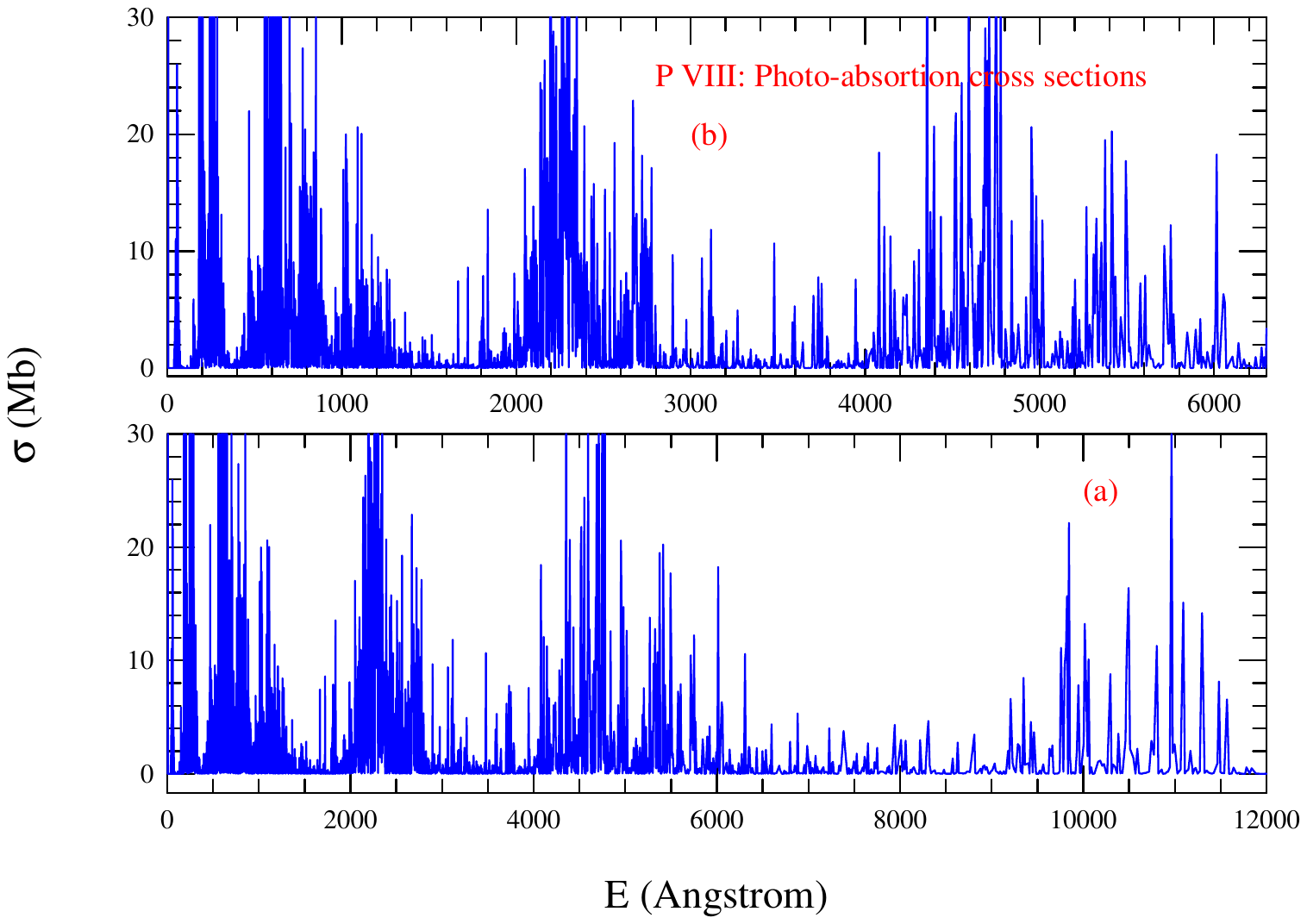}
\caption{\label{fig: p8}
The spectrum of P~VIII, created with 59219 E1 transitions. exhibiting  
strong lines in the regions of x-ray to infra-red wavelength. The 
upper panel elaborates the region of x-ray to optical. 
}
\end{center}
\end{figure*}

\subsection{P IX}

We have obtained 498 fine structure levels of P~IX. Comparison in
Table 3 shows present values are in good
agreement with those measured by Martin et al \cite{metal85}.

We have calculated 159101 transition rates, which includes 37203 E1
transitions, for N-like phosphorus, P IX. The values are compared in
Table 4 with those calculated by Cheng \cite{cheng79}, Cohen and
Dalgarno \cite{cohen64}, Naqvi \cite{cheng79}. Very good agreement 
is found among the present and those of Cheng, Cohen and Dalgarno
for the allowed E1 transitions. Present A-values for forbidden 
transitions are of comparable with order of magnitude with those by 
Naqvi \cite{naqvi51}. 

For P IX, lifetimes from two excited levels $2s^22p^3(^2P^o{1/2,3/2})$ 
from forbidden transitions, measured by Trabert \cite{trabert12}, are 
presented in Table 5. The present lifetimes are comparable but 
somewhat lower than the published values.

Photoabsorption spectrum of P~IX s presented in Fig. 9. The 
regions of strong spectral lines extends from x-ray to infra-red (lower
panel). The upper panel elaborates the region up to optical wavelengths.
\begin{figure*}
   \begin{center}
 \includegraphics[width=6.00in,height=3.50in]{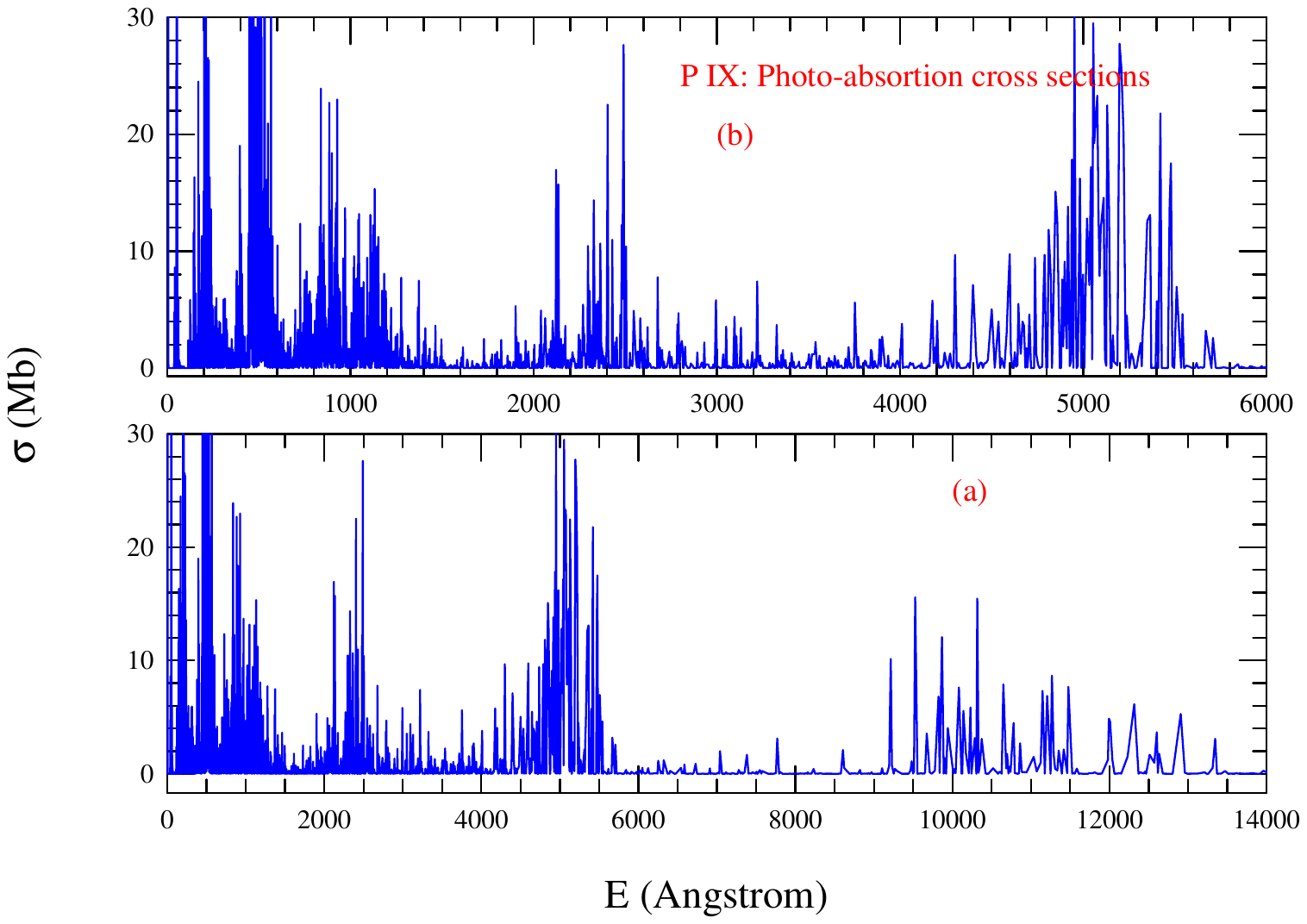}
\caption{\label{fig: p9}
The spectrum of P~IX, created with 37203 E1 transitions, is exhibiting  
strong lines in the regions of x-ray to infra-red wavelength in panel
(a). The upper panels elaborates the region of x-ray to optical. 
}
\end{center}
\end{figure*}

\subsection{P X}

We have obtained 430 fine structure levels of P~X . Comparison in
Table 3 shows good agreement with measured values.

We present a total of 113971 transitions for P~X of which 27758 are
of type E1. We present example transitions in Table 4 with comparison 
with Cheng \cite{cheng79}, Cohen and Dalgarno \cite{cohen64}
and Frose-Fischer \cite{froese66} for the allowed transitions. Very good
agreement are found among most of the transitions. Agreement of the present
values varies with the forbidden transitions with those of Naqvi \cite{naqvi51}
and Malville and Berger \cite{malv65}.  

For P X, lifetimes from 7 excited levels are presented in Table 5. There
has been more studies on this ion compared to other P ions.  Our 
values for these levels compare very well with both experimental and 
theoretical ones \cite{trabert12,trabert80a,cheng79,wiese69,nico73,faw79}
except for one odd level $^3P^o_2$ for which experimental value is much
larger than the theoretical predictions.

Photoabsorption spectrum of P-X is presented in Fig. 10. The 
regions of strong spectral lines extends from x-ray to optical but
shows some lines in the infra-red (lower
panel). The upper panel elaborates the region up to optical wavelengths.
\begin{figure*}
   \begin{center}
 \includegraphics[width=6.00in,height=3.50in]{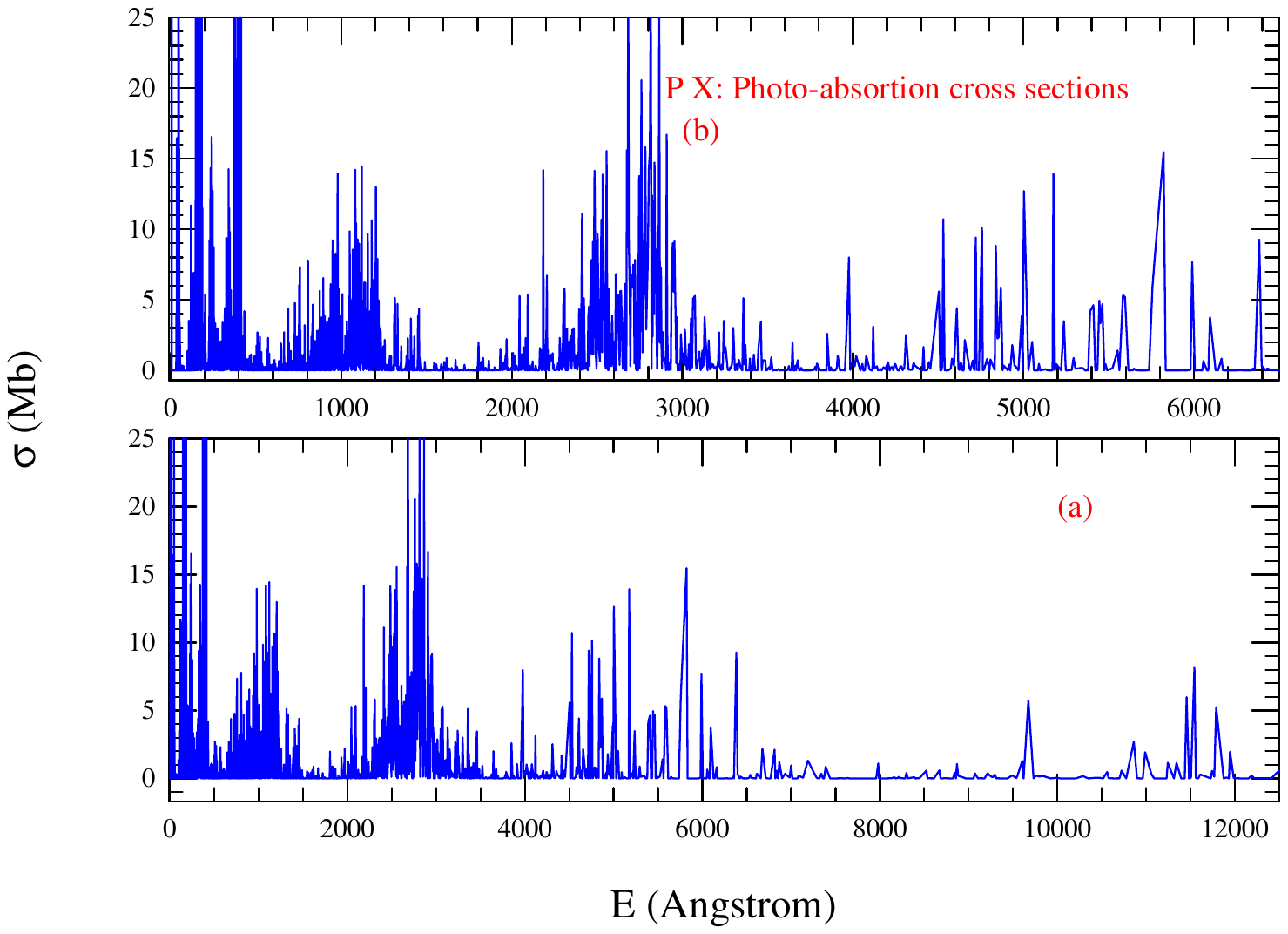}
\caption{\label{fig: p10}
The spectrum of P~X, created with 27758 E1 transitions, is exhibiting  
strong lines in the regions of x-ray to optical wavelengths in panels, 
but some in the infra-red wavelength regions. The 
upper panels elaborates the region of x-ray to optical. 
}
\end{center}
\end{figure*}

\subsection{P XI}

We have obtained 290 fine structure levels of P~XI  which are in very
good agreement with the measured values of Martin et al \cite{metal85} 
presented in Table 3.

With 290 levels, we have obtained 57130 transitions, including 14446
of them of type E1, for B-like phosphorus (P XI). We present sample 
A-values along with comparison with Cheng \cite{cheng79}, Cohen and
Dalgarno \cite{cohen64}, Naqvi \cite{naqvi51}, and Wiese et al
\cite{wiese69}. We find typical agreements, some transitions are 
very good agreement and some show differences. 

For P XI, lifetimes of 4 excited levels are presented and compared
in Table 5. General comparison of all lifetimes, theoretical and measured,
show similar ranges of values although present values are slightly lower.
Relatively large discrepancy between theory and experiment is noted 
for level $1s^22s2p^2(^2D_{3/2}$ where measured value is much lower
than the predicted values.

Photoabsorption spectrum of P~XI is presented in Fig. 11. The 
regions of strong spectral lines extends from x-ray to optical followed
by  sparse relatively weak lines.
\begin{figure*}
   \begin{center}
 \includegraphics[width=6.00in,height=2.75in]{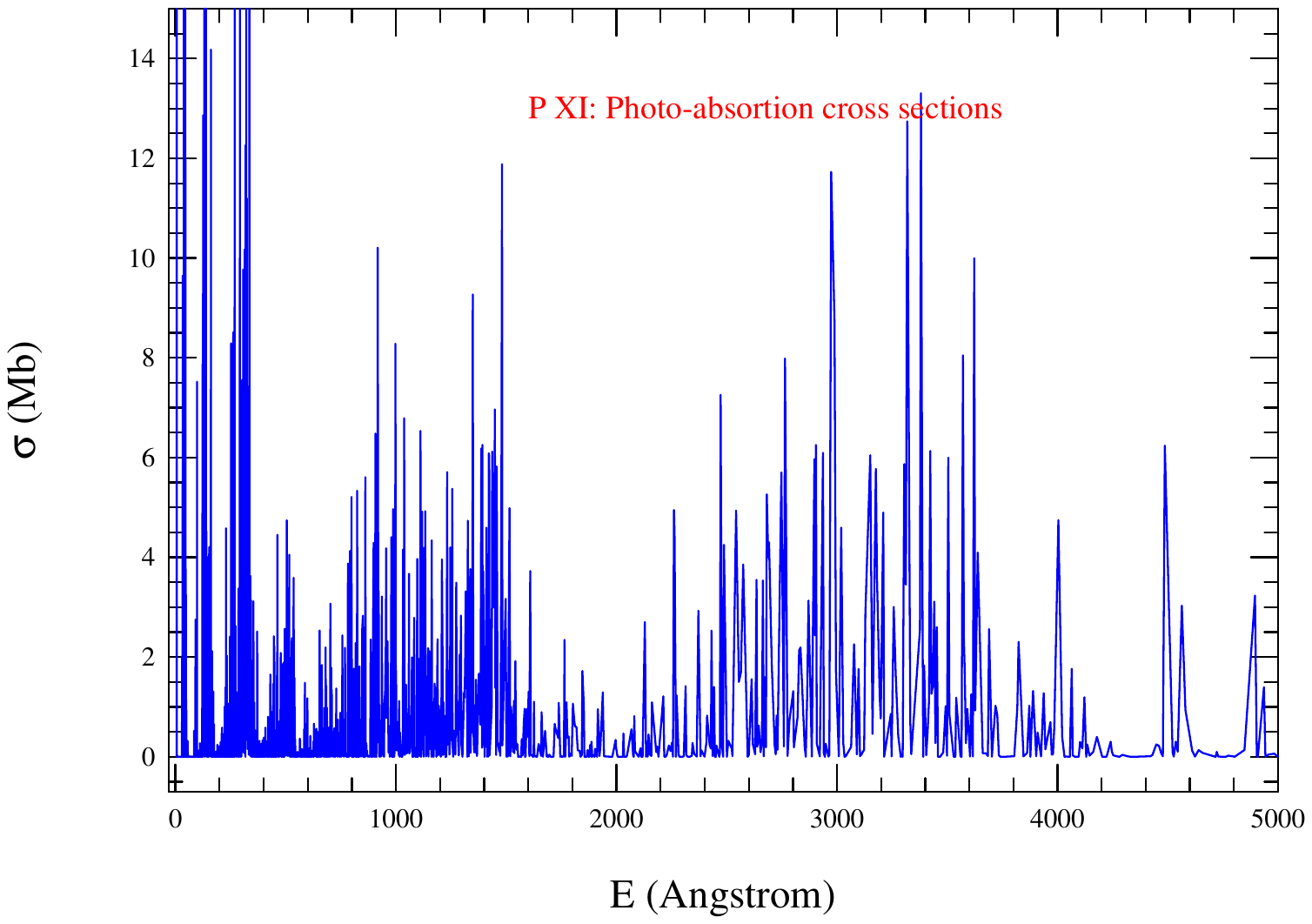}
\caption{\label{fig: p11}
The spectrum of P~XI, created with 14446 E1 transitions, is  exhibiting  
strong lines in the regions panel of x-ray to optical.
}
\end{center}
\end{figure*}

\subsection{P XII}

We have obtained 191 fine structure levels of P~XII  which are in 
about within 1\% agreement with the measured values of Martin 
et al \cite{metal85} presented in Table 3.

Transition rates for Be-like phosphorus (P XII) have been calculated 
for 20857 transitions including 4956 of type E1. Comparison with 
others have been made in Table 4 with Cheng \cite{cheng79}, Naqvi
\cite{naqvi51}, Garstrang and Shamey \cite{garstang67}, Cohen and 
Dalgarno \cite{cohen64}, and Naqvi and Victor \cite{naqvi64}. The 
allowed transitions agree very well with each other. Like some other 
ions, present A-values for forbidden transitions have varying degrees 
of agreement with those of Naqvi. 

For P XII, lifetimes from 4 excited levels are presented in Table 5.
For this ion, very good agreement is found between the experimental
and theoretical lifetimes for all levels.

Photoabsorption spectrum of P~XII is presented in Fig. 12. The 
regions of strong spectral lines extends from x-ray to infra-red. Beyond
19000 $\AA$, lines of moderate strength appear but sparsely. 
\begin{figure*}
   \begin{center}
 \includegraphics[width=6.00in,height=2.75in]{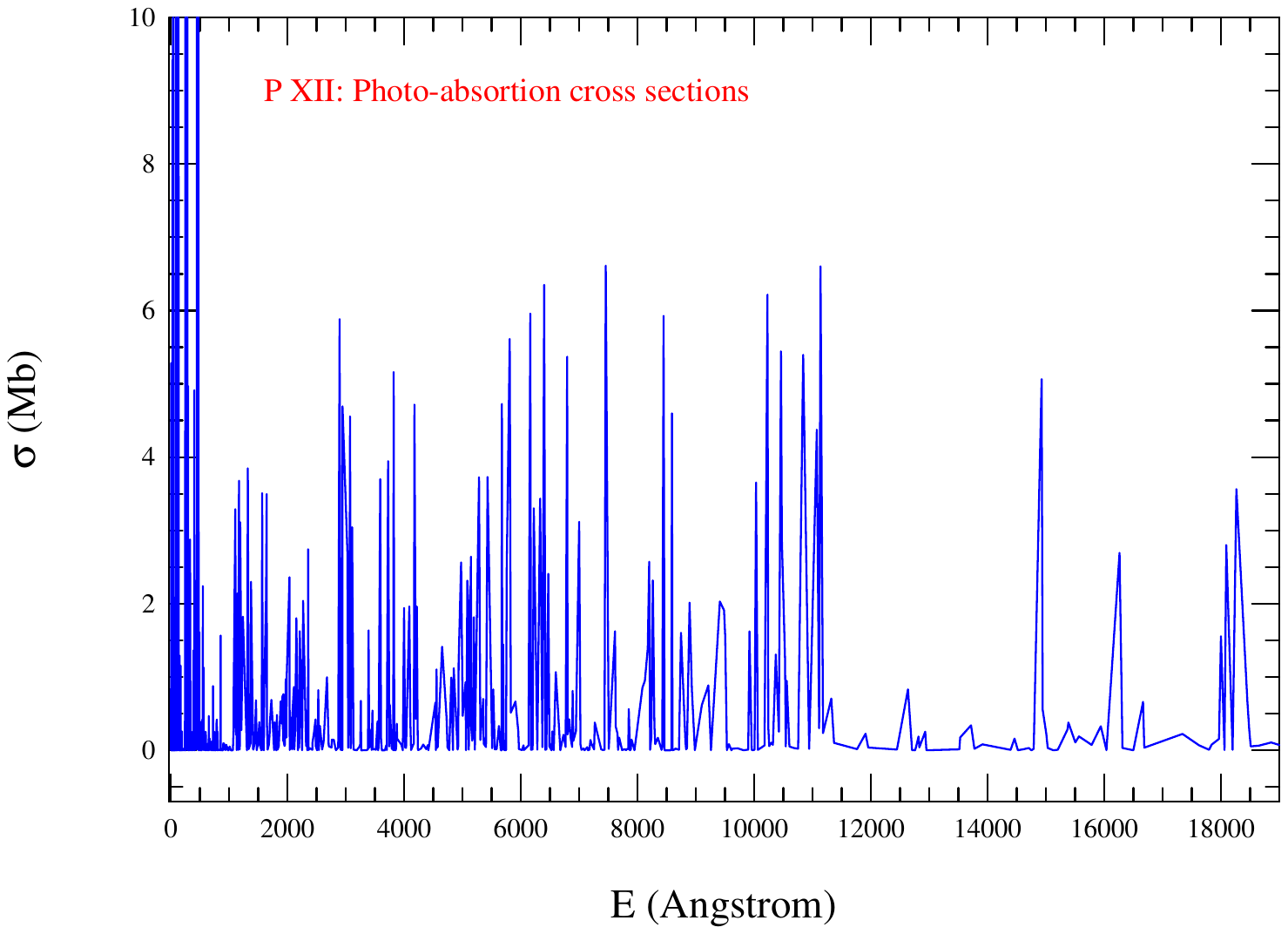}
\caption{\label{fig: p12}
The spectrum of P~XII, created with 4956 E1 transitions, is  exhibiting  
strong lines in the x-ray to infra-red regions.  }
\end{center}
\end{figure*}

\subsection{P XIII}

We have obtained 207 fine structure levels of P~XIII  which are in 
very good agreement with the measured values of Martin et al 
\cite{metal85} presented in Table 3.

We have calculated 28110 transition rates, including 6841 for E1
transitions, for Li-like phosphorus (P XIII). 
Some of them are tabulated in Table 4 for comparison with those calculated 
by \cite{zhu2016}, \cite{cheng79}, \cite{wiese69}, \cite{cohen64}
others \cite{zhu2016, cheng79}. A very good agreement exists  between 
the present and reported values. 

A single measured lifetime for P XIII, $1s2s2p(^^4P^o_{5/2}$, \cite{desc82} 
is compared in Table 5. The present value 0.162 ns is comparable to the 
measured value 1.2$\pm$0.1 ns and predicted value 1.3 ns \cite{cheng74}. 
Present value is slightly higher.

Photoabsorption spectrum of P~XIII is presented in Fig. 13. The 
regions of visible spectral lines extends from x-ray to infra-red
although becoming sparse after 13000 $\AA$. 
\begin{figure*}
\vskip 0.20in
\hskip -0.5in
 \includegraphics[width=6.00in,height=3.50in]{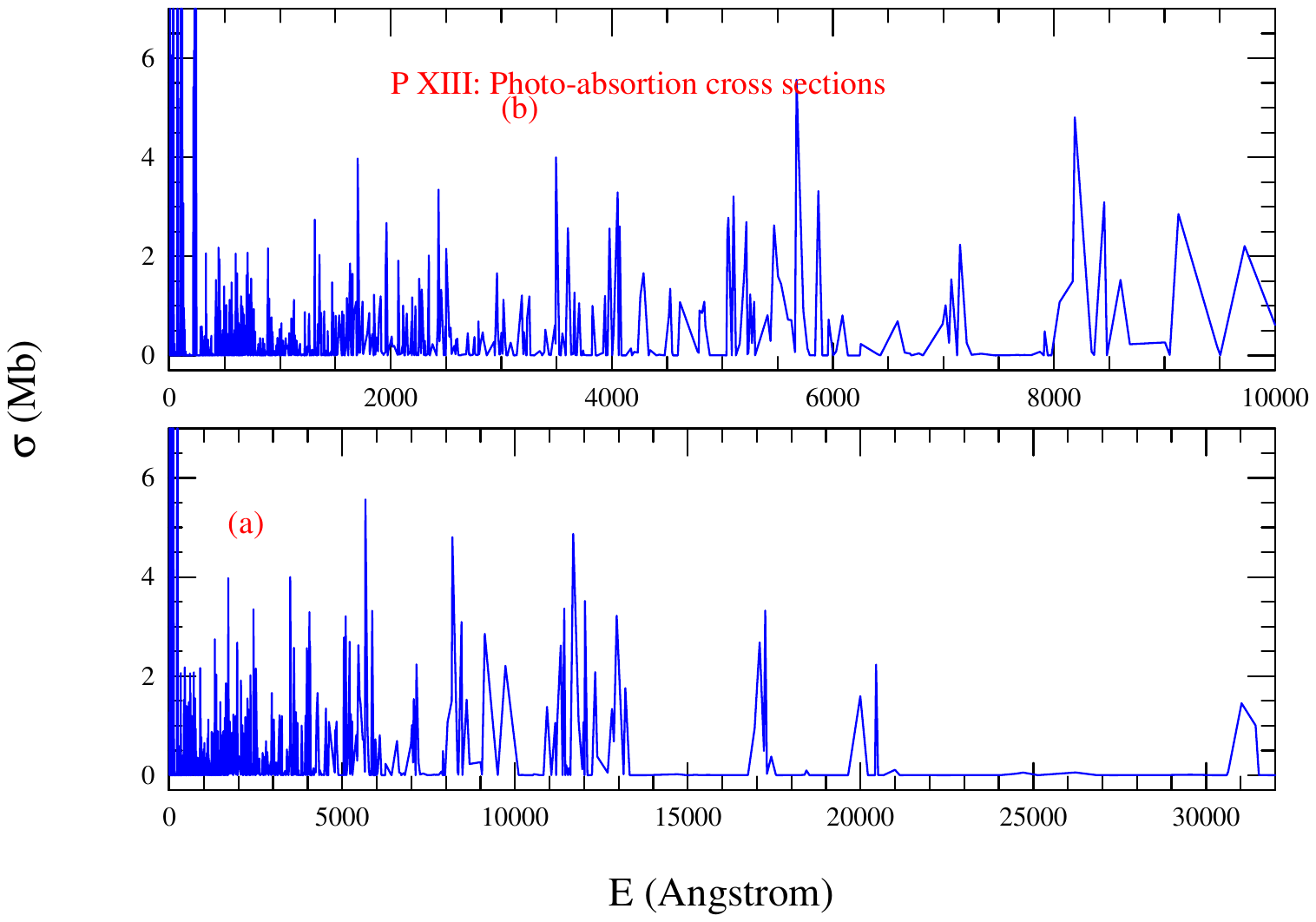}
\caption{\label{fig: p13}
The spectrum of P~XIII, created with 6841 E1 transitions, is exhibiting  
visible lines in the regions of x-ray to infra-red in panel a. 
Spectral region up to 10000 $\AA$ is elaborated in the upper panel.
}
\end{figure*}

\subsection{P XIV}

We have obtained 120 fine structure levels of P~XIII  which are in 
very good agreement with the measured values of Martin et al 
\cite{metal85} presented in Table 3.

We have obtained transition rates for He-like phosphorus (P XIV) for 
8389 transitions of which 2200 are of type E1. The present A-values are 
compared in Table 4 with those calculated by others 
\cite{zhu2016,cheng79,wiese69,cohen64,lin1977} 
with very good agreement between the present and reported values for
both allowed and forbidden transitions. 

For P XIV, lifetimes of 2 excited levels presented and compared in Table
5. Present lifetimes are slightly higher and lower for the levels compared
to the measured values and one predicted value.

Photoabsorption spectrum of P~XIV  is presented in Fig. 14. The 
regions of visible spectral lines extends from x-ray to infra-red
although becoming sparse. The relative strengths of the lines are 
weaker than other P ions.
\begin{figure*}
   \begin{center}
 \includegraphics[width=6.00in,height=2.75in]{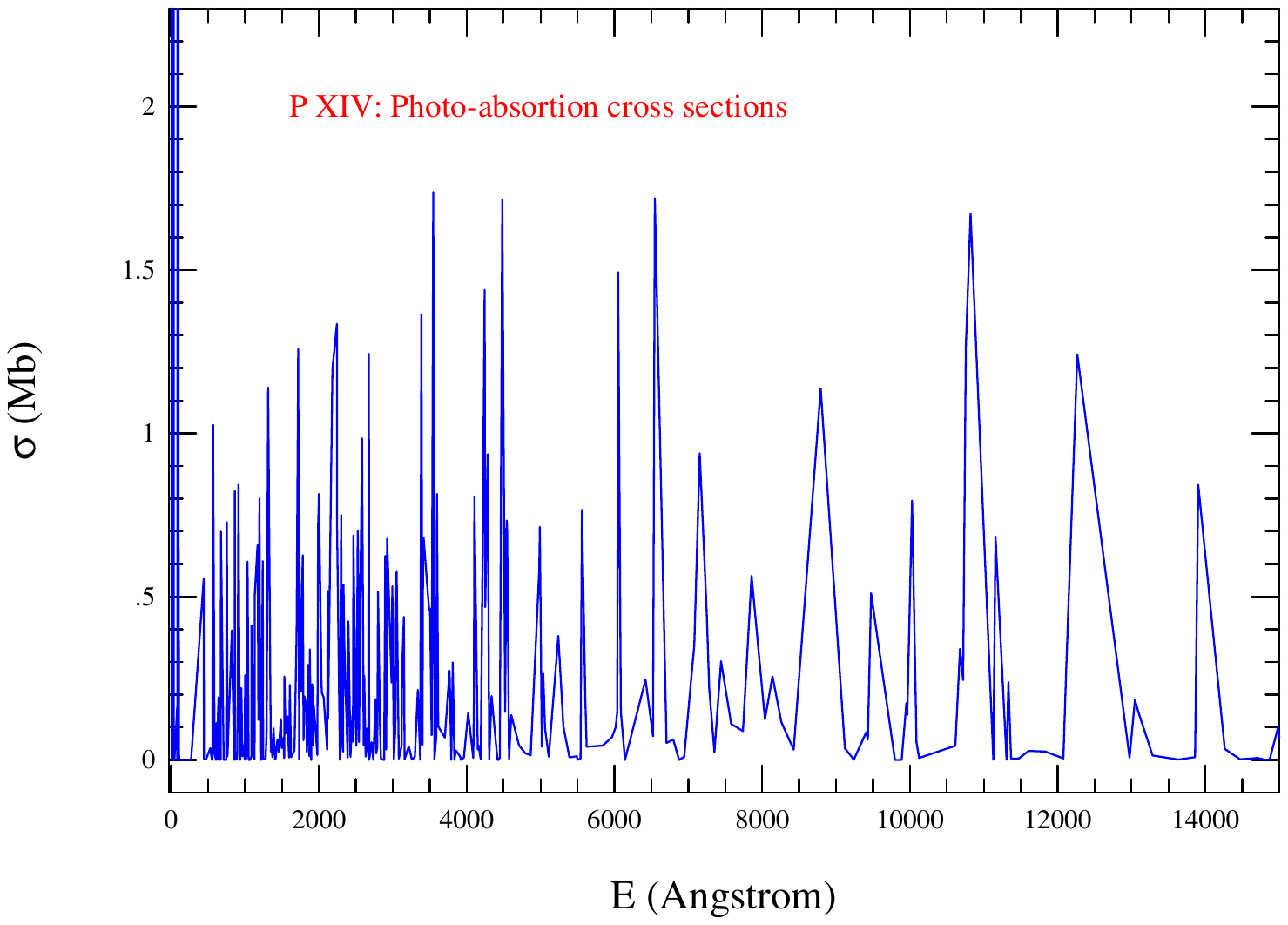}
\caption{\label{fig: 143}
The spectrum of P~XIV, created with 2200 E1 transitions, is  exhibiting  
visible lines in the regions  of x-ray to infra-red. They become sparse 
in the IR region.
}
\end{center}
\end{figure*}

\subsection{P XV}

We have obtained 16  fine structure levels of P~XV which are in less
than 1 percent difference with the measured values of Martin et 
al \cite{metal85} presented in Table 3.

We report transition rates for hydrogenic phosphorus (P XV) calculated 
for 149 transitions of which 42 are of type E1.
The values are compared with those calculated by Popov \cite{popov2017}
in Table 4. Very good agreement is found  between the present and the
available values for both allowed and forbidden transitions.

%

For P~XV, only a limited number of lines appear in the X-ray region. 
They are presented in Fig. 15.
\begin{figure*}
   \begin{center}
 \includegraphics[width=6.00in,height=2.75in]{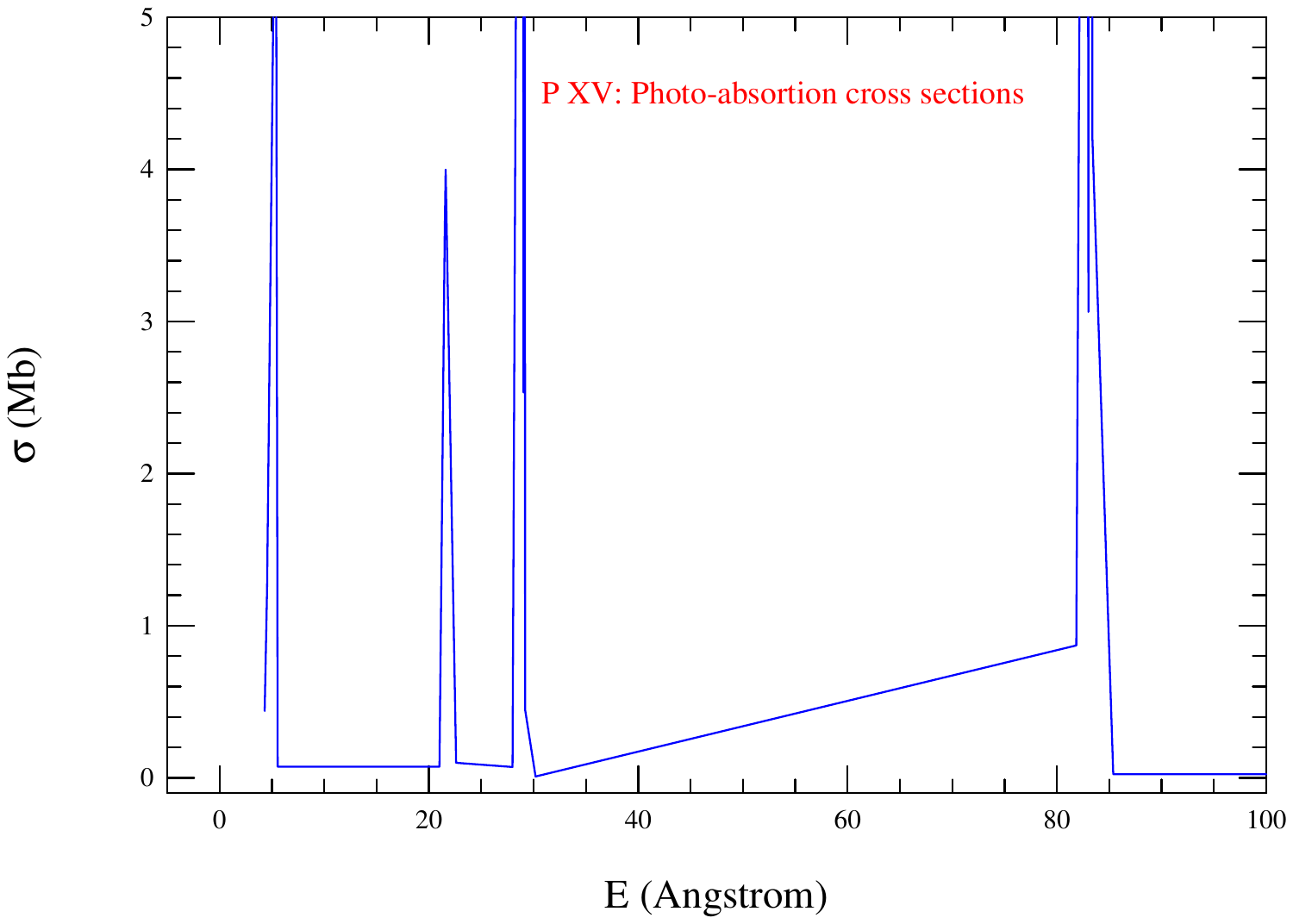}
\caption{\label{fig:p15}
The spectrum of P~XV, created with 42 E1 transitions, is  exhibiting  
a few strong lines in the x-ray region. 
}
\end{center}
\end{figure*}











\section{Conclusions}

We conclude with the following points.

i) We have carried out systematic study and present large sets of energy 
levels and transitions among them for all ionization stages of 
phosphorus, P I - P XV. The amount of atomic data obtained is much larger 
and complete than that available for most of phosphorus ions. The 
transitions are
used to produce and study the spectral features of various ionization
stages of phosphorus for the first time. 
The features can indicate presence of a P ion in an observed spectrum. 
The line strengths have been converted into photo-excitation 
cross sections to represent resonances in the synthetic spectrum of the
ion. The high peak resonances indicate higher probability of excitation 
or ionization in the spectrum. 

ii) We present energies and transitions for P~I and P~II calculated 
using Breit-Pauli R-matrix method for the first time for these ion 
and should be of high accuracy. 

iii) The present atomic data also include large sets of transition 
parameters for forbidden transitions of types E2, E3, M1, M2, 
applicable for various diagnostics.

iv) Present level energies have been compared with measured energies 
compiled in NIST Atomic Database \cite{nist} with good agreement. 
Transition probabilities from the present work have been compared 
with theoretical values reported by other investigators and, good or 
general agreement is found for each P ion. Typically the same high 
precision calculation which includes large number of levels and 
transitions can agree well to some transition probabilities and not  
well with some other transitions. The reason is that some levels 
or transitions are optimized well and some are not in the calculations. 
In such a case, general agreement should be an indication of acceptable 
accuracy.
The radiative lifetimes have been compared with experimental and 
theoretical values. We find very good agreement for some ions and 
general agreement for others.

v) We attempted for the best optimization for SS to achieve high 
accuracy results for energies, transition probabilities, and radiative 
lifetimes for each ion. The accuracy is verified and confirmed with 
comparison with existing data. Comparisons indicate that the present 
atomic data for energies, transition probabilities, and radiative 
lifetimes, and spectroscopic guidance for detection of phosphorus 
can be used for benchmarking until more accurate calculations of 
individual ions will be carried out using the R-matrix method or 
other fully relativistic approaches, which are expected to require 
significant amount of time. 

vi) The predicted spectra show considerable contributions in the 
infrared regions by P I, P II and stretching strongly toward infrared 
wavelengths by some other ions. Spectral features also indicate significant 
presence of some P-ions in the optical and UV regions. 
We have compiled atomic data and spectral features in a single 
article to study overall characteristics of the element.

\section*{Acknowledgments}

The authors are thankful to the Ohio Supercomputer Center (OSC) for 
providing time slots to carry out all computations using its High 
Performance computers.


\section*{Data Availability}

All atomic data will be available online at NORAD-Atomic-Data database
at The Ohio State University:
https://norad.astronomy.osu.edu/

\end{document}